\documentclass[10pt,journal,compsoc,usenames,dvipsnames]{IEEEtran}

\usepackage[pdftex]{graphicx}
\DeclareGraphicsExtensions{.pdf,.jpeg,.png}

\usepackage{microtype}                 
\PassOptionsToPackage{warn}{textcomp}  
\usepackage{textcomp}
\usepackage{mathptmx}                  
\usepackage{times}                     
\usepackage{cite}                      
\usepackage{tabu}                      
\usepackage{booktabs}                  
\usepackage{subcaption}
\usepackage{comment}
\usepackage{hyperref}
\usepackage{multirow}

\usepackage{tikz}
\tikzset{>=latex}

\usepackage{amsmath,amsfonts,amssymb}

\definecolor{editcolor}{RGB}{38,124,31}
\definecolor{linkcolor}{RGB}{234,23,140}
\newcommand{\revision}[1]{{#1}}
\newcommand{\linker}[1]{{\color{linkcolor}#1}}
\newcommand{\mb}{\mathbf}

\begin{document}

\title{A Generative Model for Volume Rendering}


\author{Matthew Berger,
        Jixian Li,
        and Joshua A.~Levine, \textit{Member, IEEE}
\IEEEcompsocitemizethanks{\IEEEcompsocthanksitem M. Berger, J. Li, and J. A. Levine are with the Department
of Computer Science, University of Arizona\protect\\
E-mail: \{matthew.berger, jixianli, josh\}@email.arizona.edu}}%

\markboth{Journal of \LaTeX\ Class Files,~Vol.~14, No.~8, August~2015}%
{Berger et al.: A Generative Model for Volume Rendering}

\IEEEtitleabstractindextext{%
\begin{abstract}
	\revision{
		We present a technique to synthesize and analyze volume-rendered images using generative models. We use the
		Generative Adversarial Network (GAN) framework to compute a model from a large collection of volume renderings,
		conditioned on (1) viewpoint and (2) transfer functions for opacity and color. Our approach facilitates tasks
		for volume analysis that are challenging to achieve using existing rendering techniques such as ray casting or texture-based
		methods. We show how to guide the user in transfer function editing by quantifying expected change in the output image.
		Additionally, the generative model transforms transfer functions into a view-invariant latent space specifically designed to synthesize
		volume-rendered images. We use this space directly for rendering, enabling the user to explore the space of volume-rendered images.
		As our model is independent of the choice of volume rendering process, we show how to analyze volume-rendered images produced by
		direct and global illumination lighting, for a variety of volume datasets.
	}
\end{abstract}

}

\maketitle

\IEEEdisplaynontitleabstractindextext

\IEEEpeerreviewmaketitle

\IEEEraisesectionheading{\section{Introduction}}

\IEEEPARstart{V}{olume} rendering is a cornerstone of modern scientific visualization. It is employed in a wide variety of scenarios that produce
volumetric scalar data, ranging from acquired data in medical imaging (e.g.~CT, MRI) and materials science (e.g.~crystallography), to physical simulations (e.g.~climate models and combustion).
Volume rendering offers a tool to interactively explore scalar fields, and it can be used to obtain overviews, identify distinct features, and discover interesting patterns.

In its most basic form volume rendering can be viewed as \revision{the discretization of} a physical process that models light transport through a semi-permeable material.  Specifically, given a volumetric scalar field, a viewpoint, and transfer functions (TFs)
for opacity and color, it generates an image via the volume rendering integral~\cite{max1995optical}, which governs the accumulation of
color contributions along a ray at each pixel.
Much research has been devoted to the development of TFs~\cite{kniss2002multidimensional,correa2008size,correa2009occlusion,correa2011visibility} and physically-based models 
that enhance the realism of rendered images~\cite{wald2017ospray,jonsson2017correlated}.


\revision{A user traditionally interacts with a volume renderer by modifying the TF in order to adjust optical properties in the rendered image.
In a user's workflow it is important to have tools that provide an overview of volumetric features
captured by the TF and renderer, as well as guide the user in editing the TF for further discovery of details~\cite{pfister2001transfer}. However,
traditional rendering methods such as ray casting or texture-based techniques have limitations in supporting these objectives. It is challenging to perform introspection
on a renderer in order to provide an overview of the volume. To address this, previous work has investigated sampling
the parameter space and organizing the resulting rendered images~\cite{marks1997design,jonsson2016intuitive}, or analyzing the domain space of the transfer function
to organize possible volumetric features~\cite{maciejewski2009structuring}. In addition, complexities of the rendering process present challenges in understanding
how a user's modification of input parameters impacts the output. Previous work has instead focused on analyzing the volume to understand
how changes in the data range impact selected volume features~\cite{bruckner2010isosurface,duffy2013integrating}.}

\begin{figure*}[!t]
	\begin{center}
	\begin{tikzpicture}
	\node[inner sep=0pt,anchor=south west]  at (0,0)
		{\includegraphics[width=0.9\linewidth]{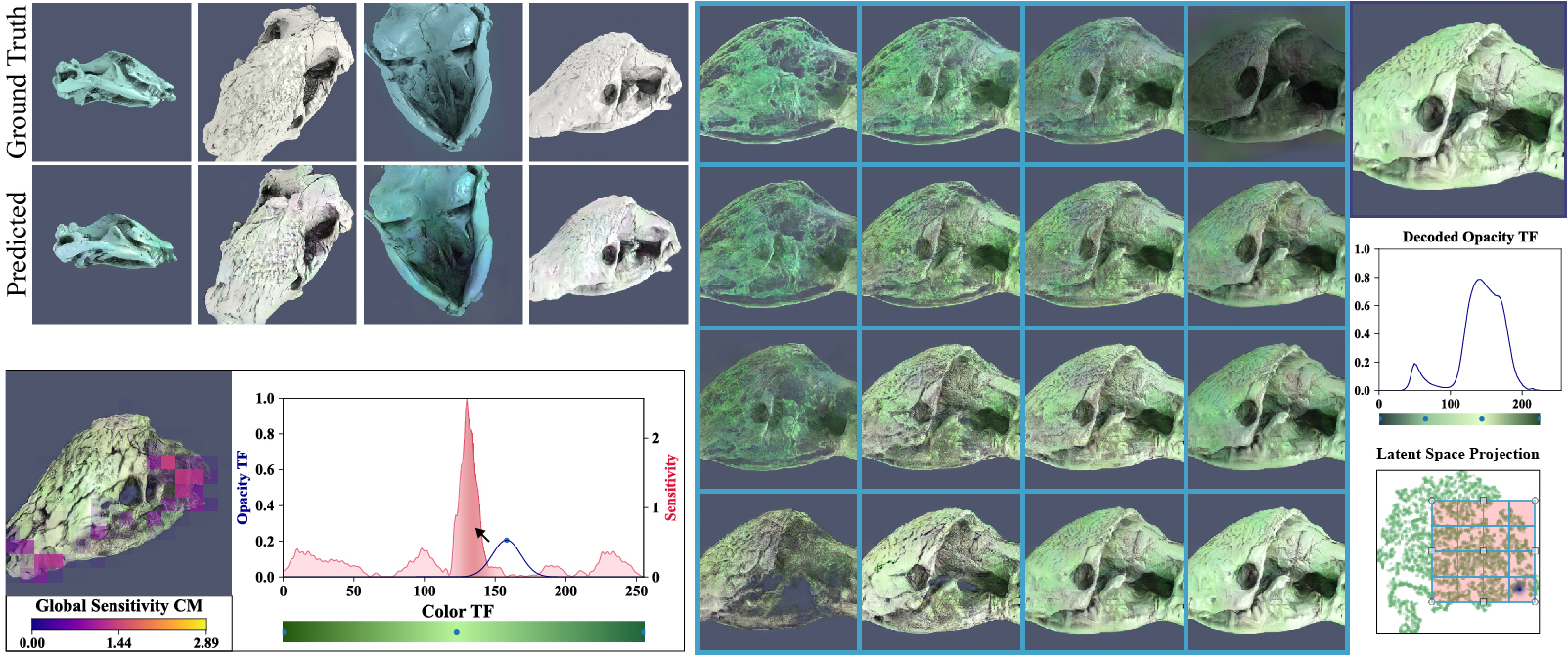}};
		\node at (3.8,3.25) {(a) Image Synthesis};
		\node at (3.5,-0.25) {(b) Transfer Function Sensitivity};
		\node at (11.8,-0.25) {(c) Transfer Function Latent Space Projection};
	\end{tikzpicture}
	  \caption{We cast volume rendering as training a deep generative model to synthesize images, conditioned on viewpoint and transfer function. In (a)
	  we show images synthesized with our model, compared to a ground truth volume renderer. Our model also enables novel ways to interact with volumetric data.
	  In (b) we show the transfer function (blue curve) augmented by a sensitivity function (red curve) that quantifies expected image change, guiding the user to
	  only edit regions of the transfer function that are impactful on the output. In (c) we show the projection of a learned transfer function latent space
		that enables the user to explore the space of transfer functions.}
	  \label{fig:teaser}
	\end{center}
\end{figure*}
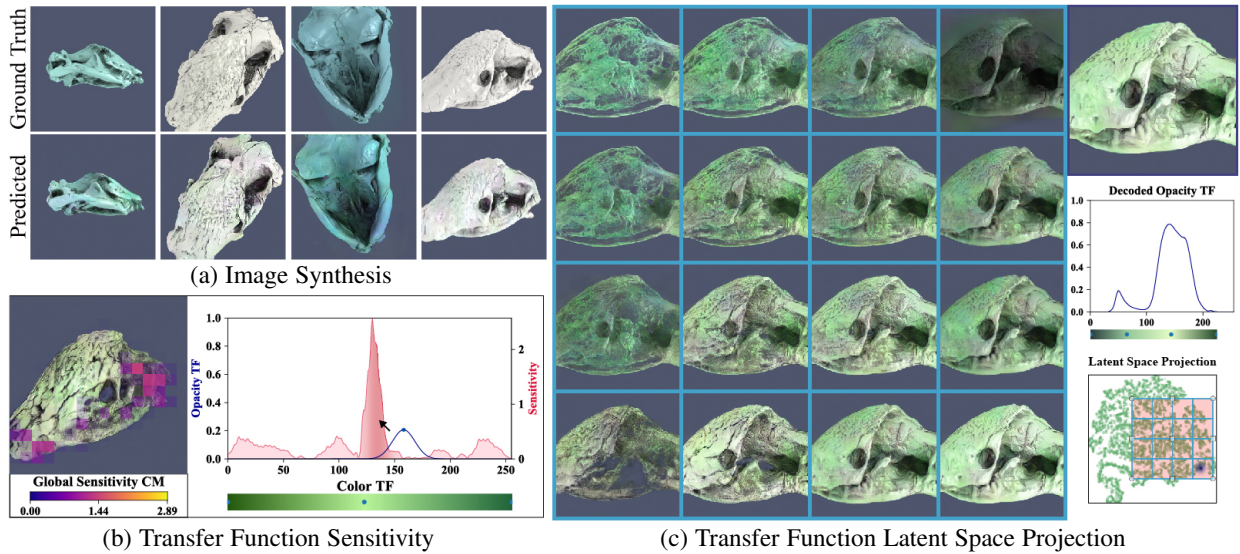

\revision{We observe that these objectives can be achieved if we consider a different way to produce volume rendered images. Instead of discretizing
a physical process, in this work we use a \emph{generative} model to synthesize renderings of a given volume.
We use Generative Adversarial Networks (GANs), a type of deep neural network which has proven effective
for representing complex data distributions~\cite{goodfellow2014generative}. In our case, the data distribution is the space of possible images produced by rendering a single volume dataset,
given a space of viewpoints and TFs (both color and opacity). The objective of the GAN is to model this distribution by training on a large collection of images.
A GAN learns this distribution by optimizing a two player game. One player is the \emph{generator}, whose job
is to synthesize samples that resemble the true data distribution as best as possible. The other player is the \emph{discriminator}, whose job is
to distinguish between samples that belong to the true data distribution from those that the generator produces.
The scenario of volume rendering presents new challenges for training GANs, due to the complex dependencies
between viewpoint, opacity TF, and color TF.
We also target images synthesized at a resolution of $256\times 256$ pixels, which pushes the practical limits of current GANs.
Our solution to these challenges is a 2-stage process tailored to volume rendering.
We first learn a GAN that generates an opacity image, conditioned on a view and opacity TF. Then, conditioned on this opacity image,
as well as the view and opacity/color TFs, we learn a second GAN that generates the final colored image.}

\revision{Our generative model is specifically designed to enhance downstream visualization applications for volume exploration, following
the analysis-by-synthesis methodology~\cite{bever2010analysis}.
More specifically, our approach computes a \emph{latent space}~\cite{hinton2006reducing} of opacity TFs that are designed to synthesize volume-rendered images, and thus captures a discriminative
space of volume features. We use this to provide the user an overview of possible volume-rendered images. We can also manipulate points in the latent space,
rather than the TF, to synthesize rendered images of the volume.
Furthermore, since our generative model is differentiable, we can compute derivatives of any differentiable function of the output image with respect
to any input parameter.
This enables us to compute TF sensitivity by taking norm derivatives of spatial regions in the output image, guiding the user towards impactful TF edits.}

\revision{Our approach is designed to complement existing volume renderers, rather than replace them. In particular, we are able to model
data distributions produced from different types of renderers.
We show the generality of our technique by modeling the distribution of volume-rendered images under basic direct illumination, in addition
to global illumination~\cite{wald2017ospray}. Thus, the benefits of a generative model for volume rendering, namely volume exploration
and user guidance, can be realized for various types of illumination. Our code is available at \linker{\url{https://github.com/matthewberger/tfgan}}, and
we summarize our contributions:}
\begin{itemize}
	\item We experimentally show the effectiveness of our technique in synthesizing volume-rendered images without explicit reference to the volume. In Fig.~\ref{fig:teaser}a
		\revision{we show the quality of synthesized images compared to ground truth renderings in the \emph{Spathorhynchus fossorium} dataset.}
	\item Pixel-level derivatives enable the notion of \emph{transfer function sensitivity}, see Fig.~\ref{fig:teaser}b. These sensitivities measure how modifications in the TF lead to changes
		in the resulting image, helping to guide the user in interactively adjusting regions of the TF based on expected change to the image.
	\item Our latent space encodes how a TF affects what is visibly rendered. This allows a user to explore the distribution of
		possible volume-rendered images without directly specifying a TF, as shown in Fig.~\ref{fig:teaser}c.
\end{itemize}



\section{Related Work}

\subsection{Volume Rendering}

Research in volume rendering spans a wide variety of areas.
We review the most relevant areas to our approach: TF design, TF exploration, \revision{compressed volume rendering,} and
applications of machine learning to volume rendering.

Transfer function design is a significant component of volume rendering, as it enables the user to interact with the volume
in finding relevant features -- see~\cite{ljung2016state} for a recent survey.
Earlier work focused on TFs defined on multidimensional histograms such as the joint distribution of scalar values and gradient
magnitude~\cite{kniss2002multidimensional} or principal curvatures~\cite{kindlmann2003curvature}.
Size based TFs~\cite{correa2008size} derive a notion of size in the volume via scale space feature detection.
The occlusion spectrum~\cite{correa2009occlusion} uses ambient occlusion to assign a value to material occlusion in the volume,
while visibility driven TFs~\cite{correa2011visibility} use view-dependent occlusion to help refine volume exploration.

Alternative approaches to TF design have been developed to help guide the user in exploration.
Rezk-Salama et al.~\cite{salama2006high} perform principal component analysis over a collection of user-provided TFs
that enables simpler interaction tools for TF exploration. 2D TF spaces driven by projected volumetric
features~\cite{de2007design} can be used to identify distinct volumetric features, while statistical features of the volume have
also been used to design statistical TF spaces~\cite{haidacher2010volume}. Image-based techniques have also been used to
support intuitive user feedback, such as in the WYSIWYG volume exploration framework~\cite{guo2011wysiwyg} and similar methods
that fuse image and TFs~\cite{wu2007interactive}. Information theoretic techniques were explored
by Ruiz et al.~\cite{ruiz2011automatic} to create TFs based on user defined view-based distributions.

Our approach for quantifying transfer function sensitivity is similar to volumetric uncertainty approaches to visualization.
Local histograms~\cite{lundstrom2006local} enable detailed evaluation of features in the volume, and a means to compute uncertainty
with respect to certain structures. Kniss et al.~\cite{kniss2005statistically} explored uncertainty volume visualization techniques
for the discernment of multiple surface boundaries.  Uncertain isocontours~\cite{pothkow2011positional} and fuzzy volume
rendering~\cite{fout2012fuzzy} explore how to guide the user in viewing volumetric data from uncertain sources.
These approaches study sensitivity of the volume, whereas our TF sensitivity measure is
strictly based on the image and the direct relationship that the TF has on all pixels in the image.

Other approaches consider how to enable the user in exploring the potentially large space of TFs.
Design galleries~\cite{marks1997design} is an early effort in organizing the space of volume-rendered images derived
from TFs, achieved by performing multi-dimensional scaling on the volume-rendered images.
This idea was extended in~\cite{jonsson2016intuitive} by embedding the images within the view of the transfer function, to better
comprehend transfer function modifications.
Transfer function maps~\cite{guo2014transfer} perform MDS based on 1D TFs for opacity and color, volume-rendered images,
and the visibility histogram~\cite{correa2011visibility}.
\revision{Image-based features, however, are view-dependent and thus one obtains different projections as the user
changes the view.}
Isosurface similarity maps~\cite{bruckner2010isosurface} \revision{are shape-based}, and provide for an exploration of the volume via the relationship
between all possible isosurfaces.
\revision{However, it is unclear how to extend isosurface similarity maps to opacity TFs. Additionally, in all aforementioned approaches it is
not possible to generate volume renderings from their respective feature spaces. In contrast, our approach computes a view-invariant opacity TF latent space
that is \emph{generative}: we can synthesize volume-rendered images from samples in this latent space.}

\revision{Our approach is related to work in compressed volume rendering, see Balsa et al. for a recent survey of techniques~\cite{balsa2014state}. Recent
methods have considered the use of multiresolution sparse coding~\cite{gobbetti2012covra} and compressed sensing~\cite{xu2014volumetric} to form
a compressed representation of the volume that is suitable for storage and rendering on the GPU. Other work has considered how to perform volume rendering
from a small set of images using camera distortion techniques and transfer function approximations~\cite{tikhonova2010visualization},
thus removing the need of the volume altogether. Ahrens et al. renders a large collection of images in-situ, and then queries these images for rendering
at runtime~\cite{ahrens2014image}. Our approach is not focused on compressing the volume, but rather focused on compressing the volume rendering process, and
novel techniques that a generative model provides for interacting with a volume renderer.}

Much less work has been devoted to the use of machine learning for volume rendering.
Early work~\cite{he1996generation} considered the use of genetic algorithms to discover TFs based on
supervision from potential volume renderings. \revision{Multi-layer perceptrons have been used to interactively classify
material boundaries in the volume~\cite{tzeng2003novel}, while Tzeng et al. interactive learns a 1D TF based on user feedback~\cite{tzeng2005intelligent}.}
Soundararajan et al. experimentally evaluate the effectiveness of different
classification schemes for generating probabilistic TFs~\cite{soundararajan2015learning}.
These approaches are \emph{discriminative} supervised learning approaches that identify user-relevant
features in the volume, whereas our method is a \emph{generative} approach for synthesizing volume-rendered images, and
shares the philosophy of Schulz et al.~\cite{schulz2016generative} in synthesizing data for visualization applications.

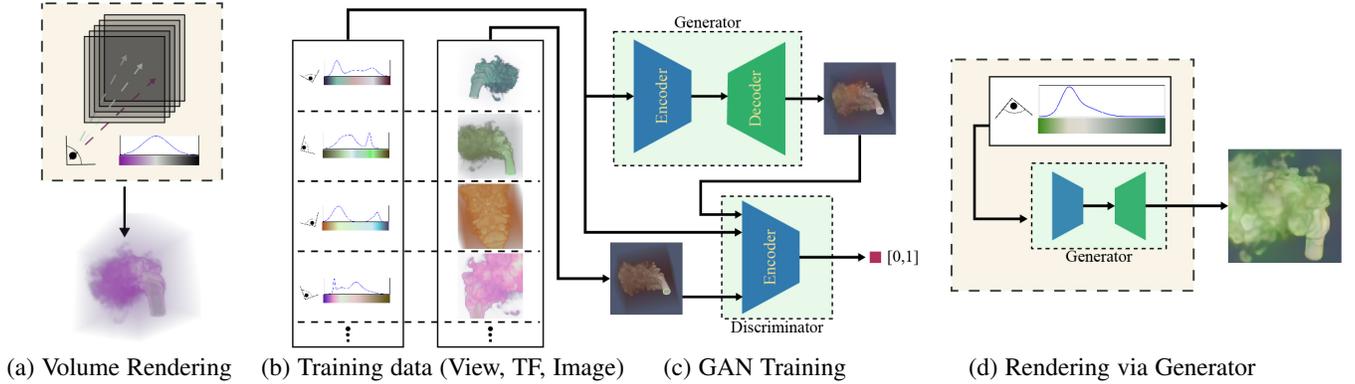
\begin{figure*}[!t]
	\begin{center}
		\begin{tikzpicture}
		\node[inner sep=0pt,anchor=south west]  at (0,0)
			{\includegraphics[width=0.95\linewidth]{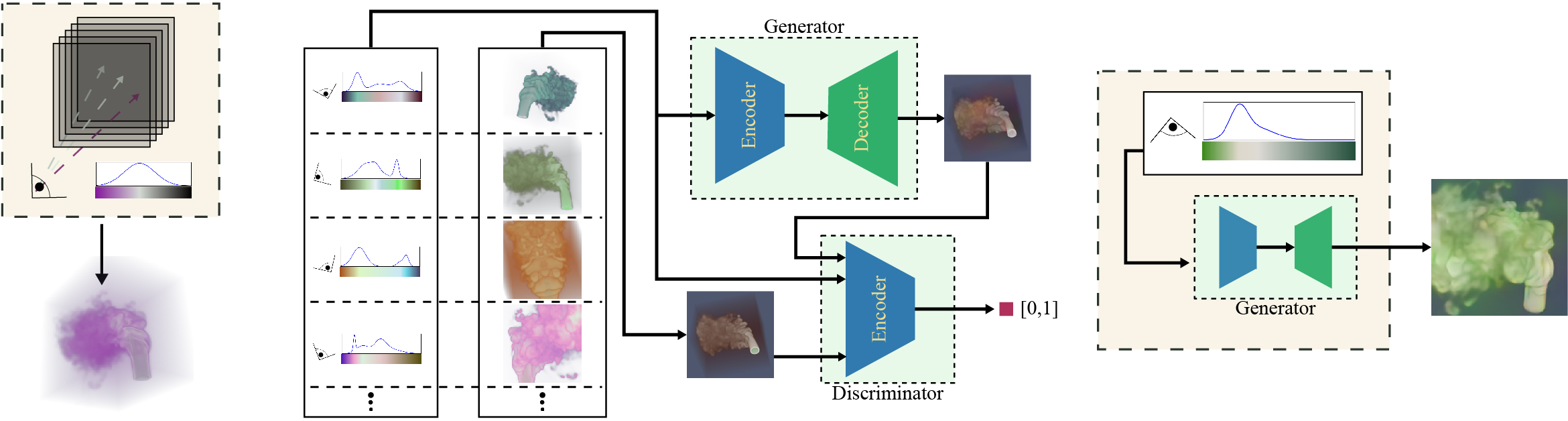}};
			\node at (1.05,-0.25) {(a) Volume Rendering};
			\node at (5.35,-0.25) {(b) Training data (View, TF, Image)};
			\node at (9.5,-0.25) {(c) GAN Training};
			\node at (14.25,-0.25) {(d) Rendering via Generator};
		\end{tikzpicture}
		\caption{ (a) Volume rendering traditionally takes as input the volume,
			viewpoint, and transfer function, and evaluates the volume rendering integral to produce an image. We interpret volume rendering as a process
			that takes just viewpoint and transfer function, and produces the corresponding volume-rendered image. We construct a generative model
			that takes a large set of volume-rendered images and (b) their visualization parameters, and (c) trains a model by learning a mapping from
			parameters to image via Generative Adversarial Networks. The trained model synthesizes images (d) from novel viewpoints and TFs,
			learning to volume render solely from viewpoint and TF.}
	\label{fig:overview}
	\end{center}
\end{figure*}

\subsection{Generative Models}

Generative models have witnessed significant advances in recent years, particular with the development of deep neural
networks. The basic idea behind generative models is to learn a data distribution from examples -- for instance,
this could be the space of all natural images. Generative adversarial networks~\cite{goodfellow2014generative} (GANs) have shown to be very effective
for generative modeling, particularly for image synthesis with complex data distributions~\cite{radford2015unsupervised,salimans2016improved}.

GANs were originally developed for generating random samples from a data distribution. It is also
possible to condition a GAN on semantic prior information, to help constrain the generation process.
This type of conditioning has been used for image generation conditioned on text descriptions~\cite{reed2016generative,zhang2017stackgan},
image inpainting via the context surrounding a missing image region~\cite{pathak2016context}, and conditioning
on a full image~\cite{Isola_2017_CVPR}. Most of these approaches condition on information which is human interpretable, and thus there
exists an expectation on the output (i.e. text describing properties of a bird~\cite{reed2016generative}). Our scenario differs from this since it is much harder
for a person to infer a volume-rendered image if only provided a TF. Rather, our work explores how GANs can
provide introspection on TFs to aid the user in volume exploration.

Our work is related to Dosovitskiy et al.~\cite{dosovitskiy2015learning}
who consider the generation of images from a class of 3D models, e.g.~chairs. They show how a deep neural network,
trained on rendered images of 3D models, can synthesize images of such renderings conditioned on viewpoint,
color, and object type (i.e.~specific type of chair). Our scenario poses unique challenges: rather than learn
from a discrete set of shapes, TFs can lead to a continuous space of shapes, and a nontrivial mapping of appearance.

\section{Approach Overview}

In order to better understand our approach, it is useful to think about volume rendering as a \emph{process} that takes
a set of inputs and outputs an image. Traditional volume rendering in its most basic form \revision{discretizes physical equations of volumetric light propagation.}
This process takes as input a volumetric scalar field,
and user-defined parameters in the form of a viewpoint and two TFs that map scalar values to opacity and color,
demonstrated in Fig.~\ref{fig:overview}a. The color of each pixel $(x,y)$ in the output image $\mb{I}$ is governed by the volume rendering integral~\cite{max1995optical}:
\begin{equation}
	\mb{I}(x,y) = \int_\mb{a}^\mb{b} \mb{c}(s) e^{-\int_a^s \kappa(u) du} ds,
	\label{eq:vri}
\end{equation}
which integrates along a ray cast from the camera position $\mb{a}$, through an image plane pixel $(x,y)$ into the volume, until it exits the volume at position $\mb{b}$. The lighting/material
contribution is captured by $\mb{c}$, while $\tau(s) = e^{-\int_a^s \kappa(u) du}$ attenuates the contribution of $\mb{c}$ as the ray travels
the space. The integral is traditionally discretized by sampling the path between $\mb{a}$ and $\mb{b}$ as a recursive compositing operation, with a user-defined $\mb{c}$ representing
the color TF -- mapping scalar value to color -- and user-defined $\tau$ representing the opacity TF -- mapping
scalar value to opacity:
\begin{align}
	\mb{I}(x,y)_{i+1} & = \mb{I}(x,y)_i + (1-\tau_i') \mb{c}_i \tau_i \\
	\tau_{i+1}' & = \tau_i' + (1-\tau_i') \tau_i,
	\label{eq:opacity}
\end{align}
where $\mb{I}(x,y)_i$ and $\tau_i$ represent the accumulated colors and opacities at each sample $i$, respectively.

We instead view volume rendering as a purely computational process: the inputs are viewpoint and TFs, and the output is
the volume rendered image, see Fig.~\ref{fig:overview}d. Note we do not make explicit use of the volume.
We instead build a \emph{generative model} by training on a large set of examples, see Fig.~\ref{fig:overview}b. Each example
is a tuple of image, viewpoint, and TFs, and the goal is to find a mapping from the viewpoint and TFs to the image, as shown in Fig.~\ref{fig:overview}c.

Given enough examples of volume-rendered examples, the learned model can then synthesize images
corresponding to novel viewpoints and TFs not seen during training, see Fig.~\ref{fig:overview}d.
Hence, the generative model can be viewed as a volume rendering engine, allowing the user to explore the space of
viewpoints and TFs even though the volume is factored out of the formulation.

This process of synthesizing images with generative models can reveal certain aspects about volume rendering, and the volume itself,
that would otherwise be challenging to capture using the volume directly and the rendering integral in Equation~\ref{eq:vri}.
First, the mapping that is learned is a subdifferentiable function with respect to the visualization parameters the user
interacts with -- viewpoint and TFs. Hence, we can compute derivatives of pixels, as well as any differentiable
function of pixels, with respect to any visualization parameter. These derivatives are used to quantify the sensitivity of TFs
to the output image, in order to guide the user in exploring distinct regions of the space of volume-rendered images.
Furthermore, the generative model can be used as a means to learn useful representations of the visualization
parameters. This is a byproduct of the model's transformation of the visualization parameters into a representation that is
more suitable for image synthesis. An analogous approach is used in prior work in image inpainting~\cite{pathak2016context}, where generative models are used to
transform an image into a more suitable representation that can be used for inpainting. In our setting volume rendering can be viewed as an auxiliary task
that, once solved, produces useful representations of visualization parameters that we use for volume exploration.

\section{Volume Rendering as a Generative Adversarial Network}

We use Generative Adversarial Networks (GANs) as our model for synthesizing volume-rendered images. In this two player game,
the generator $G$ receives as input a viewpoint and transfer function and outputs a color image $\mb{I} \in \mathbb{R}^{3 w h}$ of fixed resolution $w \times h$.
The discriminator $D$ receives as input viewpoint, transfer function, and an image, and produces a score between $0$ and $1$ indicating
whether the image is a true volume-rendering ($1$) or is a fake one produced by $G$ ($0$).
More specifically, viewpoint information is represented as $n_v$ parameters $\mb{v} \in \mathbb{R}^{n_v}$ and
TFs for opacity and color are sampled at a set of $n_t$ scalar values
yielding $\mb{t}_o \in \mathbb{R}^{n_t}$ and $\mb{t}_c \in \mathbb{R}^{3 n_t}$, corresponding to sampled versions of $\mb{c}$ and $\tau$ above, respectively.
\revision{We set $n_v=5$ corresponding to azimuth, elevation, in-plane rotation, and distance to the camera.
The azimuth angle is separated into its cosine and sine components to account for the wrap around discontinuity at $0$ and $2 \pi$.}
The TFs are uniformly sampled at a resolution of $n_t = 256$ for simplicity, though different sampling resolutions could be employed.
To simplify notation, we collectively denote the viewpoint and TFs as a single vector of visualization parameters $\mb{w}$.

\revision{The \emph{adversarial loss} in a GAN is:}
\begin{equation}
	L_{adv}(G,D) = \mathbb{E}_{\mb{I},\mb{w} \sim p_{data}} \log(D(\mb{w},\mb{I})) + \mathbb{E}_{\mb{w} \sim p_{vis}} \log(D(\mb{w},G(\mb{w}))),
	\label{eq:gan}
\end{equation}
where the first expectation is taken over the joint distribution of volume-rendered images and visualization parameters $p_{data}$, and the second is taken over the distribution of
visualization parameters $p_{vis}$.
\revision{The generator and discriminator compete in a min-max game over $L_{adv}$:}
\begin{equation}
	\min_{G} \max_{D} \,\, L_{adv}(G,D).
	\label{eq:gangame}
\end{equation}
To maximize $D$, actual volume-rendered images are predicted as real and those produced from $G$ predicted as fake.
To minimize $G$, images produced from $G$ are predicted by $D$ as real.
This game reaches an equilibrium when $D$ cannot determine real from fake,
at which point images generated by $G$ coincide with the true data distribution, i.e.~actual volume-rendered images.

\revision{We represent the generator and discriminator as \emph{deep neural networks}~\cite{goodfellow2016deep}}, due to their capability of representing
highly complex functions from large amounts of data and effective techniques for training~\cite{krizhevsky2012imagenet}.
We next discuss deep neural networks and how to utilize them for data used in volume rendering.

\subsection{Deep Neural Networks}

\revision{A deep neural network is comprised of a sequence of function compositions.}
Specifically, denoting $g_i$ as a linear function and $h_i$ as applying a nonlinear function elementwise to an input vector,
then a deep neural network is represented as an alternating sequence of linear and nonlinear functions:
$G = h_n \circ g_n \circ h_{n-1} \circ g_{n-1} \dots h_0 \circ g_0$, where a single $h_i \circ g_i$
is commonly referred to as a \emph{layer}. \revision{Each linear function has a set of parameters, and the collection of these parameters define each of the networks $G$ and $D$.
In particular, we optimize for these parameters in the solution of Equation~\ref{eq:gan}.}
We use different linear functions depending on the type of the input.

\begin{figure}[!t]
	\begin{center}
	\includegraphics[width=0.8\linewidth]{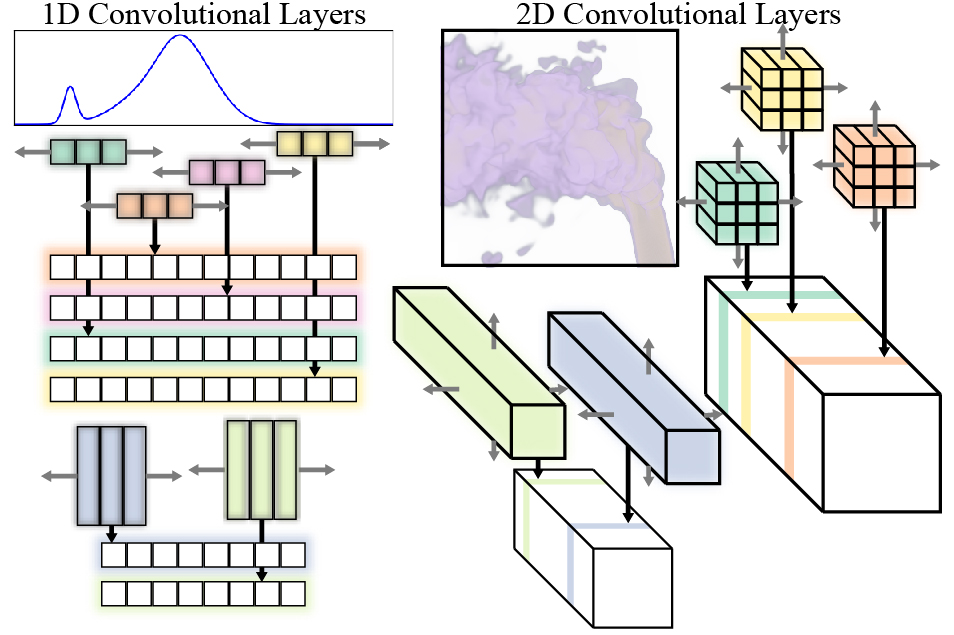}
	\end{center}
	\caption{For data with spatial dependencies, we use convolutional layers in the network. For a 1D signal on the top left,
	we show how $4$ filters convolving the signal produces a 4-channel 1D signal output. Applying $2$ filters to this then yields
	a 2-channel output. On the right, we show this for images, where the result of a 2D convolutional layer results in a multi-channel image,
	where we show $3$ filters producing a subset of channels in the output image.}
	\label{fig:pieces}
\end{figure}

\textbf{Fully Connected Layers.} Given an input of dimension $n_i$ and output dimension $n_o$,
this is a matrix $\mb{W}$ of dimension $\mathbb{R}^{n_i \times n_o}$. Namely, if $\mb{x} \in \mathbb{R}^{n_i}$ is the
output from layer $j-1$ and $\mb{z} \in \mathbb{R}^{n_o}$ is the output for layer $j$, a fully connected layer is:
\begin{equation}
	\mb{z} = h_j \circ g_j(\mb{x}) = h_j(\mb{W} \mb{x}).
\end{equation}
This is commonly used for inputs whose dimensions do not have any spatial correlation. Viewpoint information fits this case, hence
we use fully connected layers for viewpoint, following~\cite{dosovitskiy2015learning}.

\textbf{1D Convolutional Layers.} If the input has spatial dependencies, then our learned model should
respect this structure. In this case, we learn \emph{convolutional filters}~\cite{lecun1998gradient}. If the input is a set of 1D signals with spatial resolution $n_i$
containing $c_i$ \emph{channels}, or $c_i$ 1D signals each of length $n_i$, and we would like to output another set of 1D signals with $c_o$ channels, then we can define
$c_o$ filters of specified width $w$, that operate on $c_i$ input channels. Namely, if $\mb{X} \in \mathbb{R}^{n_i \times c_i}$ is the input set of 1D signals,
$\mb{Z} \in \mathbb{R}^{n_o \times c_o}$ is the target output set of 1D signals, and $\mb{W} \in \mathbb{R}^{w \times c_o \times c_i}$ are the filter weights,
then the 1D convolutional layer is defined as follows:
\begin{equation}
	Z_{a,b} = \sum_{k=1}^{c_i} \sum_{l=1}^{w} X_{s_d \cdot a + l, k} W_{l,b,k},
\end{equation}
where $s_d$ is an integer \emph{stride}, for which $s_d > 1$ results in a downsampling of the original signal, and determines
the output resolution $n_o$. Fig.~\ref{fig:pieces} (left) visually illustrates a 1D convolutional layer, where two layers are
shown. Note that unlike fully-connected layers, the filter weights do not depend on the input spatial coordinates, and
are shared across all of the input via the convolution. After a 1D convolution is performed, a nonlinearity is similarly
performed elementwise. We use 1D convolutional layers to process TFs, since these are 1D signals that contain spatial dependencies.

\textbf{2D Convolutional Layers.} This layer is very similar to 1D convolutional layers, except applied to
an image. Filters have a specified width and height, and convolution is performed in 2D, otherwise the mapping
between layers is conceptually the same as the 1D case, see Fig.~\ref{fig:pieces} (right). Strides are
similarly defined for width and height, and represent a subsampling of the image.
We also use \emph{batch normalization} in these layers~\cite{ioffe2015batch}. \revision{Batch normalization stabilizes training
by normalizing the data using mean and standard deviation statitistics computed over small amounts of data (batches).}

\textbf{Nonlinearities} \revision{Our networks primarily use two types of nonlinearities.} The generator uses Rectified Linear Units (ReLUs), defined as $h(x) = \max(0,x)$ for
element $x \in \mathbb{R}$, and the discriminator uses Leaky ReLUs, defined as $h(x) = \max(0,x) + \alpha \min(0,x)$
for parameter $\alpha$~\cite{radford2015unsupervised}.

\subsection{Network Design}

\revision{A traditional network design for GANs is the so-called DCGAN architecture~\cite{radford2015unsupervised}. Namely, $G$ transforms a given low-dimensional vector to an
image through a series of layers that interleave upsampling and 2D convolution, while $D$ transforms an image into a low-dimensional vector through a series
of 2D convolutions of stride $2$, producing a score between $[0,1]$. Pertinent to our scenario, the DCGAN can be made conditional by transforming input parameters through
$G$ to synthesize an image, while $D$ fuses image features with input parameter features~\cite{reed2016generative}. Although effective for
simple, low-dimensional inputs and small image resolutions, for instance $64 \times 64$, synthesizing volume-rendered images at $256 \times 256$ pixels presents
challenges:}
\begin{itemize}
	\item The relationship between viewpoint, opacity TF, and color TF is very complex with respect to the shape and appearance of volume-rendered images.
		\revision{Learning a transformation of these parameters for image synthesis} poses difficulties in GAN training.
	\item Generating color images of $256 \times 256$ pixels, is very difficult for GANs~\cite{salimans2016improved,zhang2017stackgan}. \revision{GAN training is unstable if
		the generator's data distribution does not overlap with the discriminator's data distribution~\cite{arjovsky2017wasserstein}, and this problem is made worse
		as the image resolution increases.}
	\item{\revision{Unlike previous GAN approaches, the generator must be designed to enable introspection on its inputs in order to help analyze volume-rendered images.}}
\end{itemize}

Inspired by previous work~\cite{wang2016generative,zhang2017stackgan}, our solution to these challenges is to break the problem down into
two simpler generation tasks, both represented as separate GANs.
The first GAN takes as input the viewpoint and opacity transfer function, and produces a $64 \times 64$ opacity image measuring only the values
produced by Equation~\ref{eq:opacity}. The opacity image captures the general shape and silhouette, as well as varying opacity in the image,
and hence is much easier to predict. \revision{In addition, we minimize an autoencoder loss with respect to the opacity TF,
in order to capture a latent TF space.}
The second GAN takes as input the viewpoint, \revision{the opacity TF's representation in the latent space}, color TF,
as well as the preceeding opacity image, to produce the final color image. Conditioning on the opacity image allows us to restrict
the regions of the image that are involved in the prediction of the final output, serving to stabilize GAN training.
\revision{Furthermore, for both generator networks the inputs -- viewpoint and TFs -- are processed independently and then merged for image synthesis. This
enables downstream analysis of the network post training.}

\subsubsection{Opacity GAN}
\label{subsec:opacitygan}

\begin{figure}[!t]
	\begin{center}
	\includegraphics[width=0.91\linewidth]{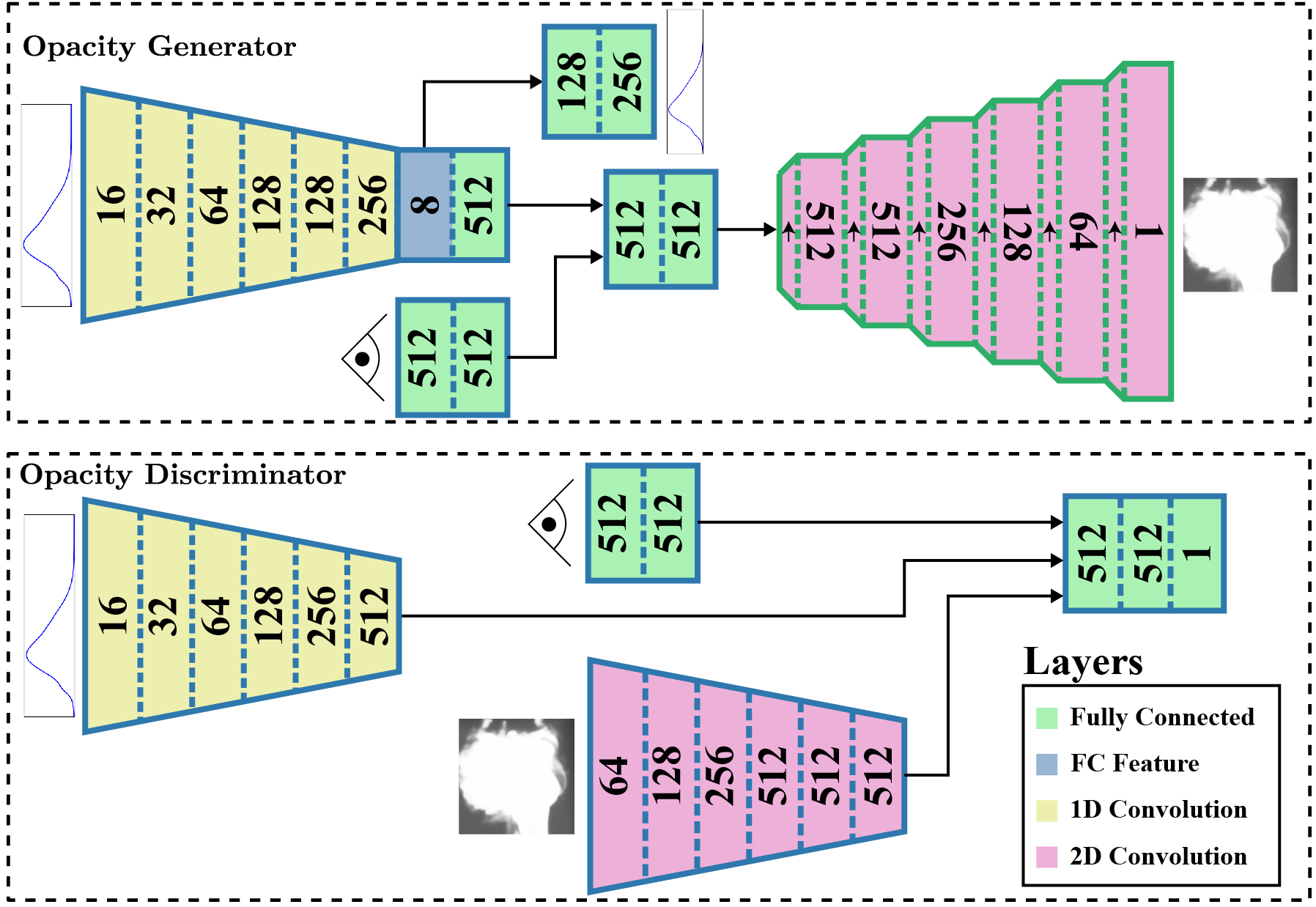}
	\caption{The architecture for our opacity GAN. Numbers indicate the feature output dimension for fully connected layers, or the number
	of channels produced in convolutional layers. 1D convolutions have width 5 / stride 2, 2D convolutions in the discriminator and generator
	have width 4 / stride 2 and width 3 / stride 1, respectively.}
	\label{fig:opnet}
	\end{center}
\end{figure}

\revision{Fig.~\ref{fig:opnet} provides network architecture details of the opacity GAN. In the generator, the opacity TF is encoded
into an 8-dimensional latent space through a series of 1D convolutions.
The encoded TF and input view are then fed through separate FC layers each producing 512-dimensional features,
and these outputs are concatenated and fed through a FC layer in order to fuse the view and TF. The fused feature} then goes through a series of interleaved upsampling
and 2D convolutional layers, using residual layers~\cite{he2016deep} to ensure well-behaved gradients, \revision{with each layer except the last using batch normalization.}
The last layer only applies a convolution, followed by a tanh activation to map the data range to $[-1,1]$, giving the final opacity image.
\revision{Additionally, we \emph{decode} the opacity TF's latent space representation through two FC layers to reconstruct the original TF.}

\revision{In the discriminator the viewpoint, opacity TF, and image are processed individually and then merged to produce a score of real/fake. Namely, the viewpoint and TF are processed
through FC and 1D convolutional layers, respectively. The image is fed through a series of 2D convolutions each of stride $2$, where each successive layer
halves the spatial resolution and increases the number of channels. The transformed viewpoint, TF, and image are concatenated and fed through a FC layer
to produce a single scalar value, followed by applying a sigmoid to produce a score between $[0,1]$.}

\revision{\textbf{Objective.} We combine the adversarial loss of Equation~\ref{eq:gan} with an autoencoder loss, ensuring that the
TF latent space is both capable of synthesizing opacity images, and reconstructing the original opacity TF:}
\begin{equation}
	\min_{G} \max_{D} \,\, L_{adv}(G,D) + \lVert G_{dec}(G_{enc}(\mb{t}_o)) - \mb{t}_o \rVert_2^2,
\end{equation}
\revision{where $G_{enc}$ and $G_{dec}$ represent the encoding of the opacity TF to the latent space, and its subsequent
decoding to the opacity TF, respectively. This ensures discriminability of the opacity TF when opacity images
for different TFs are the same, which is essential in the second stage for combining opacity and color TFs.}

\subsubsection{Opacity-to-Color Translation GAN}

\begin{figure}[!t]
	\begin{center}
	\includegraphics[width=0.91\linewidth]{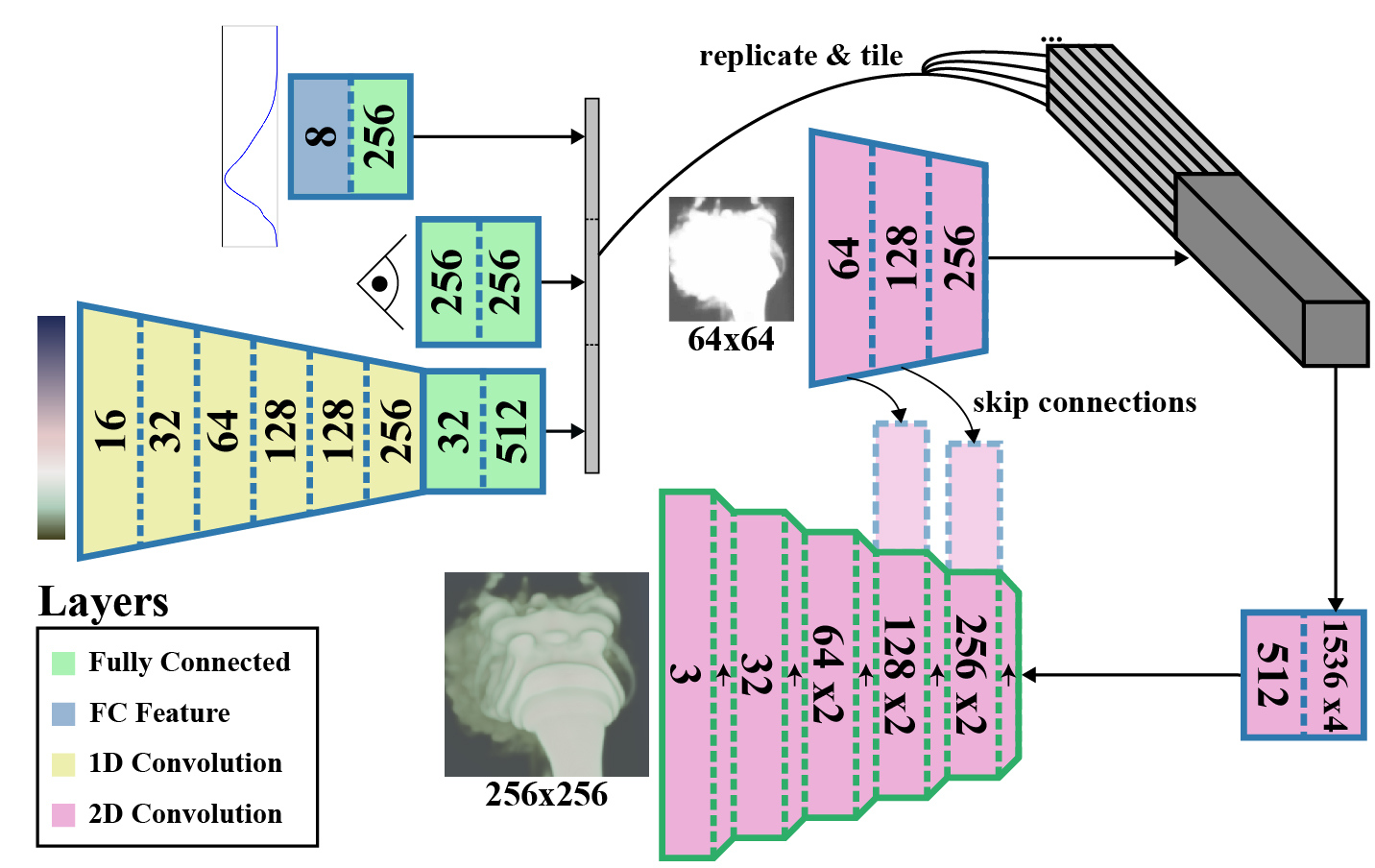}
	\caption{The generator for the opacity-to-color translation GAN, with symbols and notation consistent with Fig.~\ref{fig:opnet}.
	Skip connections, or the concatenation of the opacity image's 2D convolutional encodings onto the input of the color image's decoding,
	help enforce spatial consistency in the synthesized color image.}
	\label{fig:translate}
	\end{center}
\end{figure}

The objective of this GAN is to produce the volume-rendered $256 \times 256$ image, conditioned on viewpoint, color and opacity TFs, as well
as the $64 \times 64$ opacity image. We view this as an image-to-image translation problem~\cite{Isola_2017_CVPR}, transforming
an opacity image to a color image. Additionally, there are two factors we must consider relative to~\cite{Isola_2017_CVPR},
namely merging the opacity with the visualization parameters, and generating an image of higher resolution than the input.
We denote this the opacity-to-color translation GAN, or translation GAN for short.

The generator proceeds by transforming the viewpoint information in the same manner as the opacity GAN, while
the color TF undergoes a sequence of 1D convolutional layers, followed by a FC layer. \revision{We transform the opacity TF through
the encoder of the opacity GAN's generator (the blue layer in Fig.~\ref{fig:opnet} and~\ref{fig:translate}), and then feed this through a FC layer.
This links the opacity TF latent space between the networks, a property that we utilize in Sec.~\ref{sec:tffeats}.}
The opacity image is transformed in a similar manner as the opacity image in the opacity GAN's discriminator, but only going up
to an $8 \times 8$ spatial resolution. We then concatenate all of the visualization features, followed by replicating and tiling this
as additional channels onto the transformed image. This is then fed through a series of residual layers~\cite{he2016deep} to fuse
the image and visualization features, similar to previous work~\cite{Sangkloy_2017_CVPR,zhang2017stackgan}.

In \revision{synthesizing the $256 \times 256$} color image,
we employ \emph{skip connections}~\cite{Isola_2017_CVPR}. That is, we concatenate the outputs from each \revision{convolutional layer} of the
opacity image onto the \revision{convolutional layers of the output synthesized image}, restricted to corresponding spatial resolutions (see Fig.~\ref{fig:translate}).
Skip connections ensure that the \revision{output convolutional} layers retain the spatial structure of the opacity convolutional layers, hence we can preserve
the overall shape inherent in the opacity image. Upon producing a $64 \times 64$ image, we no longer have skip
connections from the opacity image, so we employ standard upsampling/convolution to reach the $256 \times 256$ image. These upsampling
steps serve to effectively fill in details that the low-resolution opacity image may not have captured.

The discriminator is very similar to the Opacity GAN's discriminator, the main addition being the inclusion of the color TF transformation.
We do not make use of the opacity image in the discriminator as we did not find it to provide much more discriminatory
power than just the final color image.

\revision{\textbf{Objective.} Solely using an adversarial loss for the translation GAN has several limitations.
First, we find that a good color mapping is challenging to learn, despite shape details being preserved. Furthermore, for images computed with advanced illumination
models we find that training can be unstable. To address these issues we supplement the adversarial loss with an image-based loss,
namely the $l_1$ norm difference between the ground truth image and generated image, as this has shown to be effective in addressing
the aforementioned issues~\cite{Isola_2017_CVPR,Sangkloy_2017_CVPR}. Thus, our objective for the translation GAN is formulated as follows:}
\begin{equation}
	\min_{G} \max_{D} \,\, L_{adv}(G,D) + \lambda \lVert G(\mb{v},\mb{t}_o,\mb{t}_c) - I \rVert_1,
	\label{eq:translationgan}
\end{equation}
\revision{where $I$ represents the ground truth image associated with view $\mb{v}$, opacity TF $\mb{t}_o$, and color TF $\mb{t}_c$,
and $\lambda$ weights the importance of the $l_1$ loss. In practice we find $\lambda=150$ preserves color and stabilizes training
without overly blurring generated images.}

\subsection{Training}

Each GAN is trained to optimize the min-max game of Equation~\ref{eq:gangame} with \emph{minibatch stochastic gradient descent}.
This iterative process repeatedly queries a small batch of data and the gradient of the loss function
is computed on this batch with respect to the network parameters. The parameters are then updated from the gradient, where we
use ADAM optimization~\cite{kingma2014adam}.
The gradient is constructed using \emph{backpropagation}~\cite{rumelhart1988learning}, which computes derivatives of the network by applying the
chain rule layer-wise, starting from the loss, and working back to the inputs of the network.

GANs, more specifically, are trained by alternating gradient descent steps in the discriminator and generator. First, the discriminator
updates its parameters by using a batch of real images and visualization parameters, and minimizes a binary cross-entropy loss that encourages the
discriminator to predict these images as real. Next, the visualization parameters (and opacity image in the case of the translation GAN)
are pushed through the generator to synthesize images. The discriminator is then updated to encourage a prediction of false for these
images. Last, the generator's parameters are updated by tricking the discriminator: encouraging it to predict these images as being real.

\subsubsection{Training Data}

We generate training data set by performing volume rendering over a wide range of viewpoints and TFs.
For each training data instance the viewpoint is randomly generated \revision{and the opacity TF is generated} by sampling from
a Gaussian mixture model (GMM). More specifically, we first randomly sample the number of modes in the GMM (from 1 to 5), and then
for each mode we generate a Gaussian with a random mean and standard deviation -- relative to the range of the scalar field -- and a random amplitude. 
For certain volumes there may exist scalar values that either result in a rendering where the whole volume is opaque or is nearly empty. In these
cases we manually adjust the minimum and maximum scalar values the mean values may take on, as we find the bounds of the scalar field are where
this tends to occur. The color TF is based on the opacity TF by first sampling random colors at the opacity TF GMM means and the scalar value
global extrema, \revision{and is generated by performing piecewise linear interpolation between the colors}.
We bias colors to have a higher lightness component at the means, and a low lightness at the global extrema. Correlation between
high values in the opacity TF and high lightness in the color is meant to mimic a user's intent in emphasizing volumetric features.

We note that this approach is relatively data-independent. More sophisticated semi-automatic transfer
function design techniques could be employed~\cite{maciejewski2009structuring,ruiz2011automatic} in order to limit the space, particularly if the user has prior
knowledge about the data that could guide the process. Our goal is to show the generality of our technique, and thus we
impose as few limitations as possible on the space of possible volume renderings. This is done to generalize
to as many TFs as possible, and enable interaction in an open exploration, similar to how a user would interact
with a traditional TF editor.

\section{Applications}
\label{sec:applications}

\revision{Our generative model enhances volume exploration through analysis of the space of volume-rendered images. We introduce two
applications that take advantage of the generative capabilities: transfer function sensitivity and exploration of volume rendering through
the opacity TF latent space.}

\subsection{Transfer Function Sensitivity}
\label{sec:tfsens}

Recall that our generative model is differentiable.
Thus, we can compute derivatives of pixels with respect to the TF.
The derivative of a pixel with respect to a scalar value of the TF can be used as a
way to measure \emph{transfer function sensitivity}, or to quantify how much this pixel will change if we adjust the transfer function
at the given scalar value.

\begin{figure}[!t]
	\begin{center}
		\includegraphics[width=1.0\linewidth]{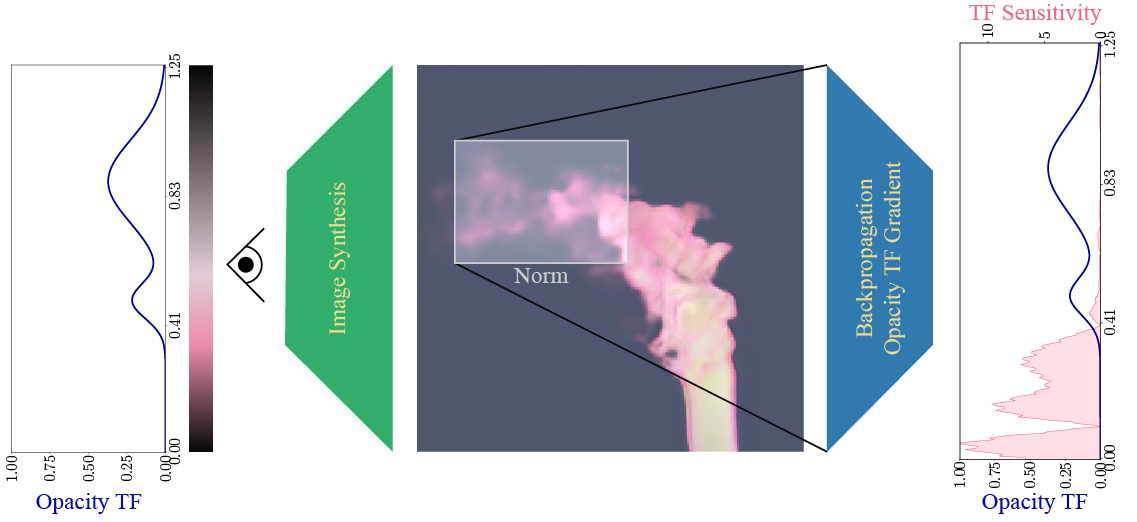}
		\caption{We illustrate the computation of opacity TF sensitivity. The input parameters are pushed through the network to obtain an image,
		then the $l_2$ norm of a user-specified image region is computed, and last the opacity TF gradient is obtained by backpropagation.}
		\label{fig:explain_sensitivity}
	\end{center}
\end{figure}

More specifically, transfer function sensitivity follows from a first-order Taylor series expansion for a given pixel in the image $I_{(x,y)}$.
Given a small additive perturbation $\delta$ of a given scalar value in a TF $a$, fixing all other visualization parameters we have:
\begin{equation}
	|I_{(x,y)}(a+\delta) - I_{(x,y)}(a)| = |\frac{\partial I_{(x,y)}}{\partial a} \delta| + O(\delta),
\end{equation}
where $O(\delta)$ are higher-order terms. Hence the partial derivative gives us a measure of expected difference in pixel value.
Note that we may also compute derivatives for any differentiable function of a set of arbitrary image pixels. In particular,
we use the $l_2$-norm of pixels for a given set of image locations $R$ as our function, and restrict sensitivity to the opacity TF $\mb{t}_o$,
since this impacts the overall shape of the volume rendering. Denoting $G_{o}$ and $G_{t}$ as the opacity and translation GANs, respectively,
transfer function sensitivity $\sigma : R \rightarrow \mathbb{R}^{256}$ is taken as the following function:
\begin{equation}
	\sigma(R) = \nabla_{\mb{t}_o} \| G_{t}((G_{o}(\mb{v},\mb{t}_o)), \mb{v}, \mb{t}_o, \mb{t}_c) \|_{R},
	\label{eq:tfsensitivity}
\end{equation}
where the $R$ subscript denotes computing the norm the set of pixels in $R$.

\revision{Fig.~\ref{fig:explain_sensitivity} illustrates the computation involved, where the image is first produced by feeding the input
parameters through the network, followed by computing the $l_2$ norm of a region $R$, and then performing backpropagation~\cite{rumelhart1988learning} to
compute the opacity TF gradient. Note that a traditional volume renderer faces difficulties in computing the TF gradient, as
it is necessary to differentiate the compositing operation in Equation~\ref{eq:opacity}, and is made worse when considering
complex illumination factors such as ambient occlusion. We use TF sensitivity to guide the user in TF editing through two complementary visualization techniques:
Region Sensitivity Plots and Scalar Value Sensitivity Plots.}

\subsubsection{Region Sensitivity Plots}

\revision{TF sensitivity is used to show where modifications in the opacity TF domain will result in large changes in the resulting output image.
This is achieved by superimposing the TF sensitivity $\sigma$ on top of the opacity TF, which we term the Region Sensitivity Plot. Namely,
since the range of $\sigma$ is the 256 scalar values in the opacity TF discretization, we plot $\sigma$ directly with the opacity TF in order to guide the user
as they interact with the opacity TF. A large value of $\sigma$ suggests a large change in the output. The user can specify a region in the image $R$, and we interactively
update the Region Sensitivity Plot based on $R$ in order to guide the user in their TF edits. The right-hand side of Fig.~\ref{fig:explain_sensitivity}
shows an example Region Sensitivity Plot for a user-specified region.}

\subsubsection{Scalar Value Sensitivity Fields}

\revision{We also use TF sensitivity to construct a scalar field over the image domain. The field is the TF sensitivity defined over image regions, conditioned on
a scalar value, which we call the Scalar Value Sensitivity Field. Specifically, we first define a grid resolution $r$ and divide the image into
$r \times r$ blocks. For each block we then compute the TF sensitivity in Equation~\ref{eq:tfsensitivity}. This produces a 3-tensor
$\Sigma \in \mathbb{R}^{256 \times r \times r}$, where $\Sigma(i,\cdot,\cdot) \in \mathbb{R}^{r \times r}$ is a field defined on the $r \times r$
image blocks for the scalar value at index $i$. Setting $r = 256$ computes sensitivity for each pixel, however this is prohibitively costly to perform,
as it requires performing backpropagation $256^2$ times. Thus we set $r$ to 8 or 16 depending on acceptable latency for the user.
We accelerate computation by performing backpropagation in parallel on the GPU over minibatches of size 64.}

\revision{This set of scalar fields is useful in understanding what parts of the image are likely to change,
based on modifying certain ranges of the opacity TF. This complements Region Sensitivity Plots:
Scalar Value Sensitivity shows sensitivity over the image conditioned on a scalar value in the opacity TF domain, whereas Region Sensitivity shows sensitivity
in the opacity TF conditioned on an image region.
We combine both techniques into a single interface, as shown in Fig.~\ref{fig:teaser}(b). We plot TF sensitivity with
respect to the entire image, and show Scalar Value Sensitivity as the user hovers over the TF domain. The user thus obtains an overview
of scalar values expected to result in changes to the image, and by hovering over the TF domain they observe details on where
in the image changes are likely to occur. Since a user's TF edit tends to impact a localized range of scalar values, we anticipate this in
visualizing the field by applying a Gaussian filter to
the sequence of fields centered at a selected scalar value for a given bandwidth, where the filter weight for each Scalar Value Sensitivity Field is
superimposed on the Sensitivity Plot in red. In order to provide global and local context, we color map sensitivity based on the global range of the field, and encode the range specific to
a user's selection via opacity mapping.}

\begin{figure}[!t]
	\begin{center}
		\includegraphics[width=1.0\linewidth]{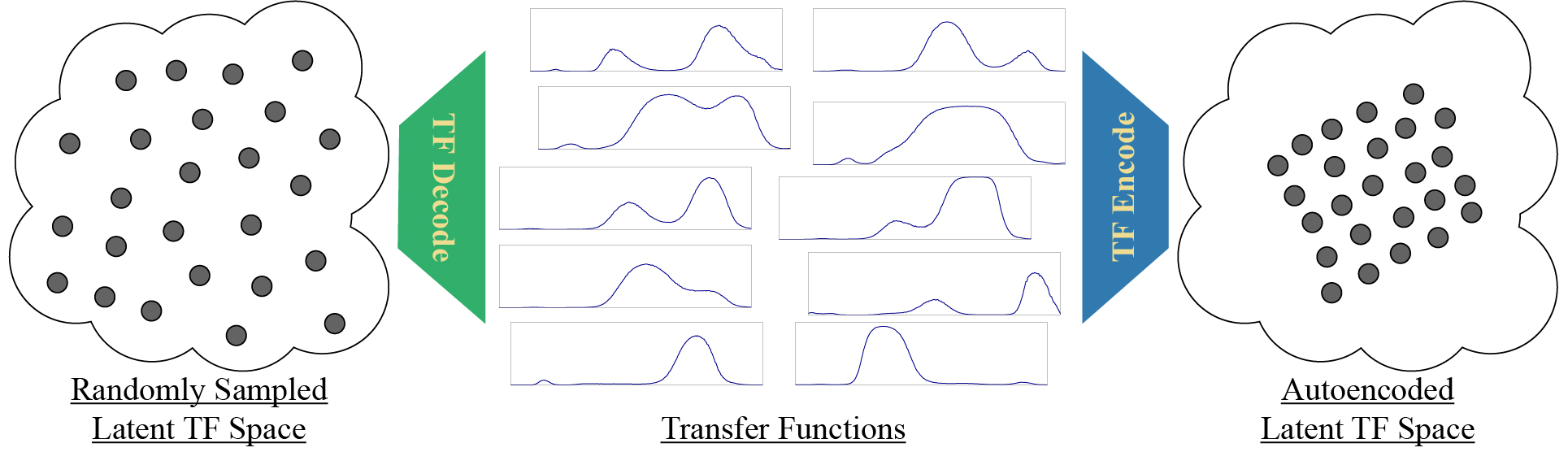}
		\caption{The opacity TF latent space is sampled by first performing uniform sampling, decoding each sample
		to reconstruct a TF, and then encoding the set of TFs back into the latent space.}
		\label{fig:explain_latent}
	\end{center}
\end{figure}

\subsection{Exploring the Opacity TF Latent Space}
\label{sec:tffeats}

\revision{A byproduct of the generative model in its synthesis of volume-rendered images is the network's encoding of visualization parameters.
Recall that the opacity TF is transformed into an 8-dimensional latent space through the opacity GAN, from which we synthesize the opacity image and reconstruct the original TF.
This dimensionality reduction forces the network to learn a latent space that is informative. Specifically, the latent space must capture all possible variations
of \emph{shape} admitted by the opacity TF in a manner that is also \emph{predictive} of the original TF. We use the latent space to provide the user an exploration
of all possible features present in the volume. We achieve this through four steps: sampling the latent space, projecting points in the latent space to 2D,
structured browsing of the latent space, and opacity TF latent space interpolation for detailed inspection.}

\begin{table*}[!t]
\begin{center}
   \begin{tabular}{| c | c | c | c | c | c || c | c |}
      \hline
		Dataset & Resolution & Precision & Size (MB) & Rendering Model & Training Images Creation & Image RMSE & Color EMD \\ \hline
		\multirow{3}{*}{Combustion} & \multirow{3}{*}{$170 \times 160 \times 140$} & \multirow{3}{*}{float} &\multirow{3}{*}{15} & No Illumination & 2.7 hours & 0.046 & 0.011 \\ \cline{5-8}
		& & & & Direct Illumination & 5 hours & 0.060 & 0.011 \\ \cline{5-8}
		& & & & Global Illumination & 14 hours & 0.060 & 0.010 \\ \hline
		Engine & $256 \times 256 \times 110$ & byte & 7 & No Illumination & 3 hours & 0.061 & 0.015 \\ \hline
		Visible Male & $128 \times 256 \times 256$ & byte & 8 & Global Illumination & 14 hours & 0.075 & 0.013 \\ \hline
		Foot & $256 \times 256 \times 256$ & byte & 16 & No Illumination & 3.3 hours & 0.064 & 0.017 \\ \hline
		Jet & $768 \times 336 \times 512$ & float & 504 & No Illumination & 4 hours & 0.086 & 0.022 \\ \hline
		Spathorynchus & $1024 \times 1024 \times 750$ & byte & 750 & Global Illumination & 5 days & 0.116 & 0.020 \\ \hline
   \end{tabular}
\end{center}
\caption{We show dataset characteristics on the left and quantitative evaluation of our model on held-out test sets on the right.}
\label{table:datasets}
\end{table*}

\textbf{Sampling the Latent Space.} \revision{Not every point in the latent space corresponds to a valid TF and opacity image. It is necessary to first
discover a subspace, or more generally submanifold, of the latent space on which valid TFs exist. To this end, we use the decoder in our TF autoencoder as a means
of sampling TFs. We first sample points in the latent space uniformly at random, in our experiments $10^4$ samples, and then push the samples through the decoder to obtain a set of TFs.
We then transform these TFs back to the latent space via the set of 1D convolutional layers in our opacity GAN's generator, see Fig.~\ref{fig:explain_latent}. This process effectively probes the range
of the TF decoder, producing TFs similar to those seen during training. In practice, we find that the decoder is not injective for points in the latent space that do not correspond
to valid TFs. Experimentally, we find that encoding the set of decoded TFs results in latent space samples that have low-dimensional structure,
observed by computing the Singular Value Decomposition of the samples and finding a falloff in the singular values.}

\textbf{2D Projection.} \revision{We next take the set of samples in the latent space and project them into 2D. We use t-SNE~\cite{maaten2008visualizing}
in order to best preserve geometric structure in the samples. We use a perplexity of $30$ in our experiments in order to not bias the perception of clusters in the data.
Fig.~\ref{fig:teaser}(c -- lower right) shows an example t-SNE projection for the \emph{Spathorhynchus fossorium} volume.}

\textbf{Structured Latent Space Browsing.} \revision{In order to enable an overview of the volume, we structure the latent space by allowing the user to brush a $4 \times 4$ rectangular grid on the
2D projection, and synthesize an image for each grid cell given the cell's set of contained opacity TF latent space samples.
More specifically, for a given grid cell we compute the mean of this set of samples and synthesize the image from the mean, alongside the view and color TF.
For efficiency, we push the $4 \times 4$ set of inputs through the network in a single minibatch, enabling interactivity for manipulating and viewing the grid of images.
In Fig.~\ref{fig:teaser}(c -- lower left)
we show an example grid layout of images given a user's selection in the 2D projection (lower right), depicting the major shape variations in the volume. As the user selects
smaller rectangular regions, finer grained variations can be viewed in the resulting image grid, since the set of points to average in each cell will cover a smaller
portion of the latent space.}

\begin{figure}[!t]
	\begin{center}
	\includegraphics[width=0.8\linewidth]{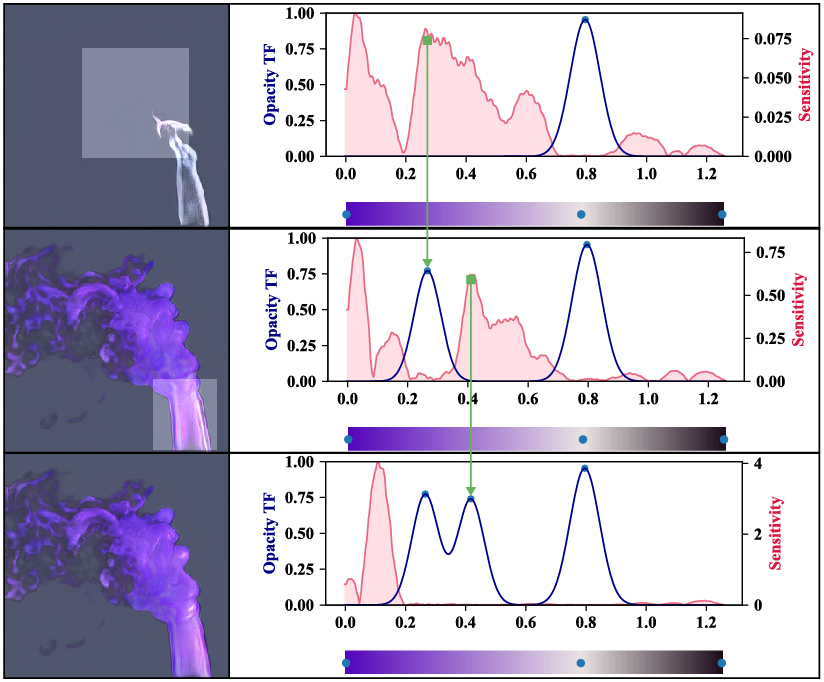}
		\caption{Region-based sensitivity helps to drive a user's opacity TF edits. Upon selecting a region, the user
		observes the sensitivity plot, and then can select modes to add in the opacity TF that suggest large change in the image.}
	\label{fig:sense_region}
	\end{center}
\end{figure}

\textbf{Latent Space Interpolation.} \revision{We also allow the user to select specific regions in the latent space projection for more detailed inspection. For a given point in the 2D projection,
highlighted in blue in Fig.~\ref{fig:teaser}(c -- lower right), we perform scattered data interpolation of latent opacity TFs for all points located in a disk centered at the selected point.
We use Shepard interpolation with a Gaussian kernel whose bandwidth is $\frac{1}{3}$ of the specified radius, taken as $5\%$ of the 2D bounding box diagonal.
The synthesized image is shown in Fig.~\ref{fig:teaser}(c -- upper right), in addition to the reconstructed TF shown in the middle right, corresponding to the TF decoded from the
interpolated latent TF. Thus, the user can gain an understanding of the space of TFs as they explore the projection.}

\section{Experimental Results}

We demonstrate the quality and uses of our model in several ways. \revision{We first show applications of TF sensitivity
and the TF latent space projection for exploring volume rendering. We then validate our network through quantitative
and qualitative evaluation, and study parameter choices in our model.}

\begin{figure}[!t]
	\begin{center}
	\includegraphics[width=0.8\linewidth]{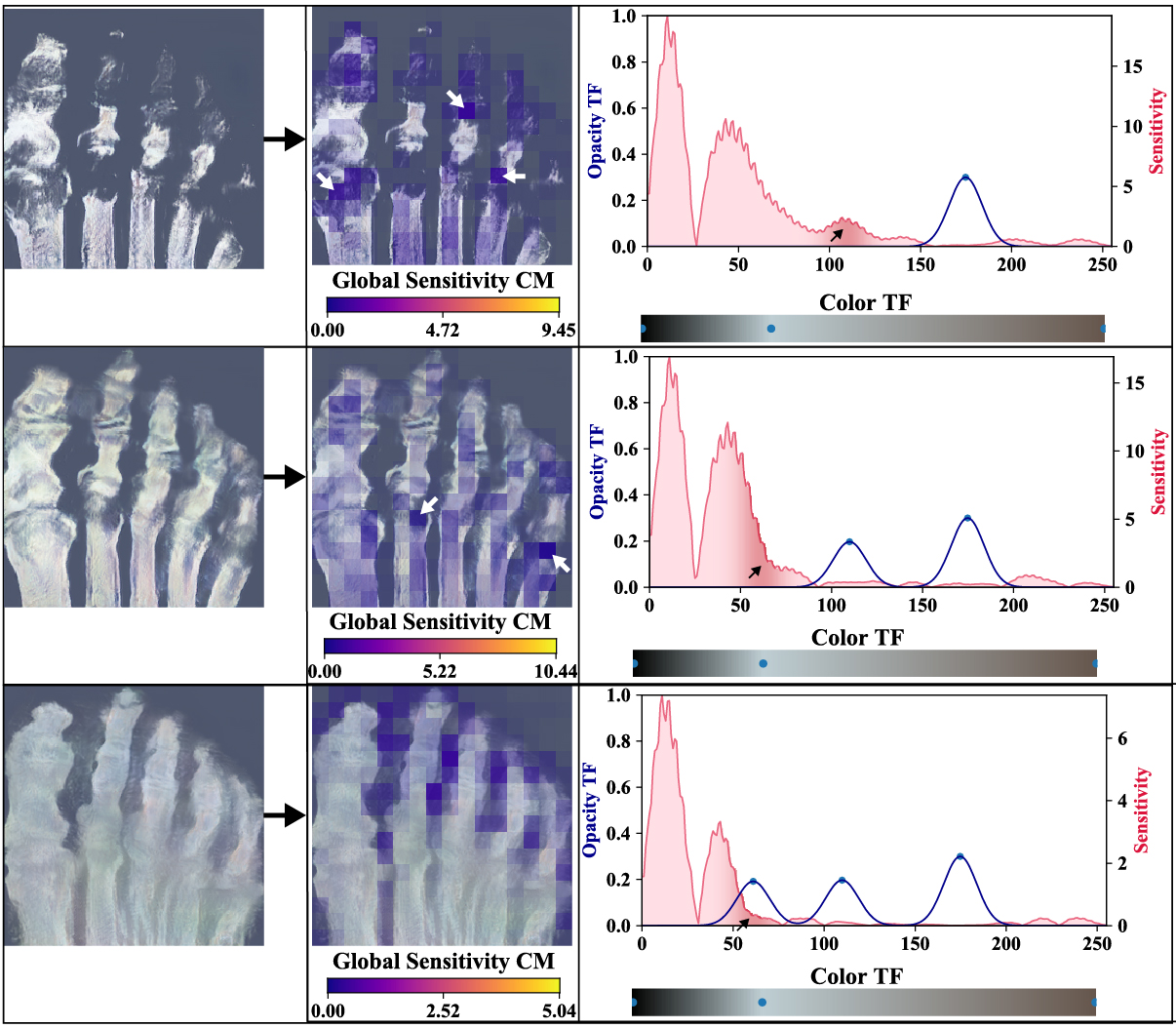}
		\caption{The Scalar Value Sensitivity Field enables the user to visualize image regions that are likely to change, given a user selection
		in the opacity TF domain. This helps the user modify TF values that correspond to changes in spatial regions of interest.}
	\label{fig:sense_field}
	\end{center}
\end{figure}


\begin{figure*}[!t]
	\begin{center}
	\includegraphics[width=1\linewidth]{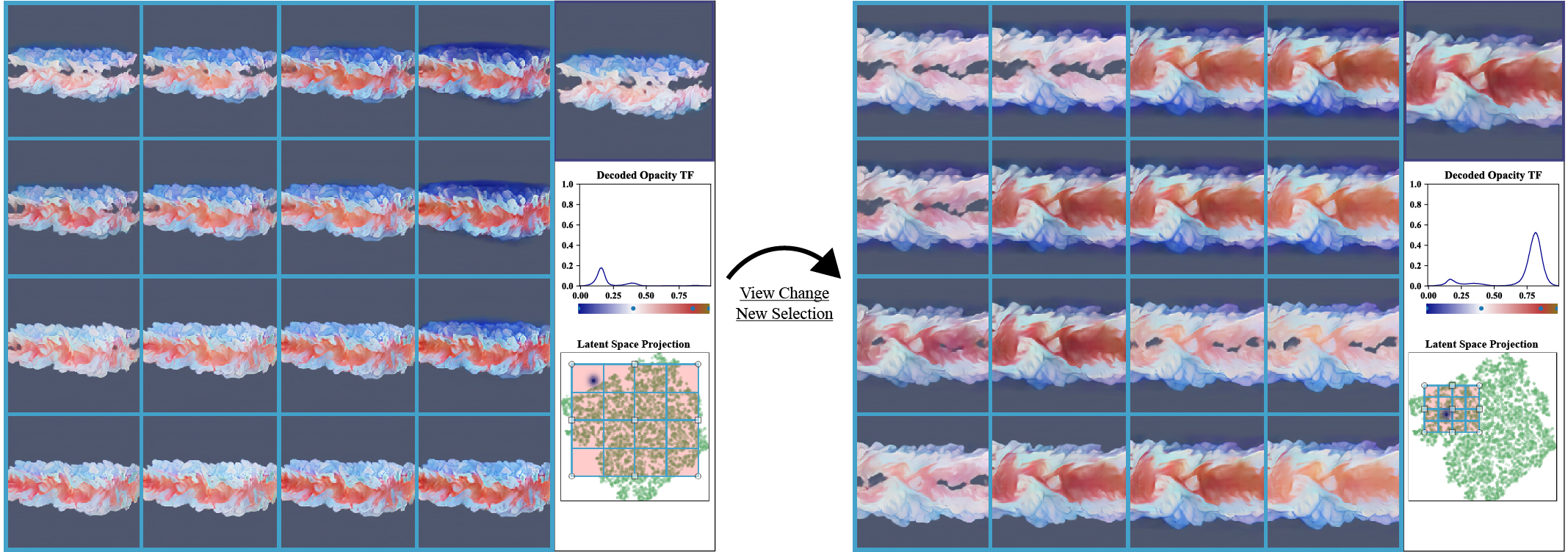}
		\caption{We show opacity TF exploration through 2D projection of the TF latent space sampling. On the left the user selects most of the projection in order
		to obtain an overview of volumetric features, while still enabling details through direct selection in the projection, shown as the blue Gaussian blob that corresponds
		to the upper right image and reconstructed TF in the middle right. Selection of a smaller region on the right enables the study of finer-grained shape variation.}
	\label{fig:exploration}
	\end{center}
\end{figure*}

\begin{figure}[!t]
	\begin{center}
		\includegraphics[width=0.9\linewidth]{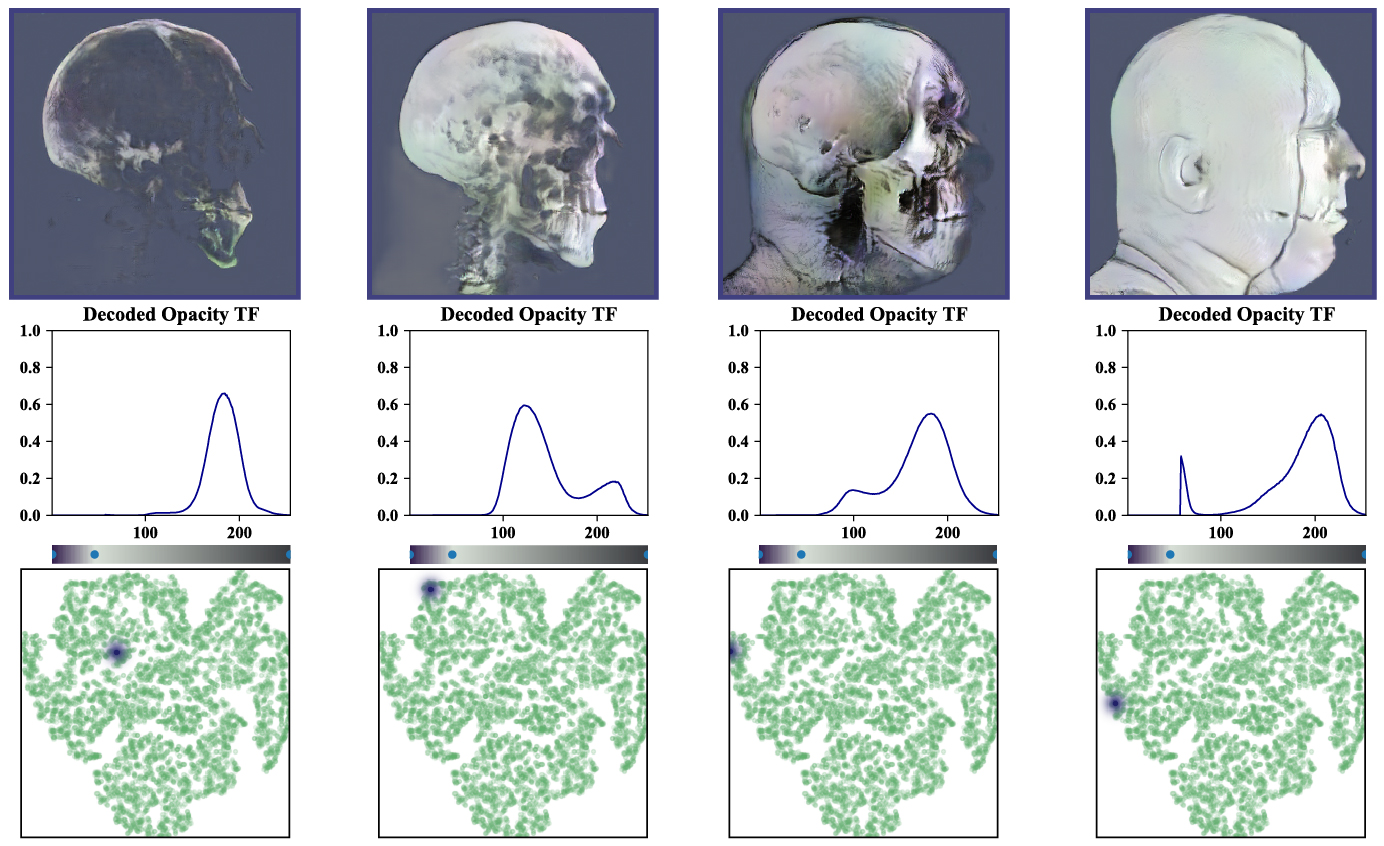}
			\caption{A user's browsing through the projected latent TF space (bottom) can aid in their understanding of the space of opacity
			TFs (middle) based on the synthesized images (top).}
		\label{fig:tf_interpolate}
	\end{center}
\end{figure}

\textbf{Implementation Details}. We have implemented our network in PyTorch~\footnote{\url{http://pytorch.org}}, using an NVIDIA GTX 1080 Ti GPU
for all network training and experiments. In training the opacity GAN we set the learning rate to $2 \times 10^{-4}$, and halve it every $5$
\emph{epochs}, or passes over the dataset. For the translation GAN the learning rate is set to $8 \times 10^{-5}$, and halved every $8$ epochs.
The color TF is represented in L*a*b color space.
We use minibatch sizes of 64 and 50 for the opacity and translation GANs, respectively. The training data size for each experiment is 200,000 samples.

\textbf{Datasets}. \revision{Our experiments use the following volume datasets: a Combustion simulation dataset, the Engine block, Visible Male,
and Foot datasets~\footnote{\url{http://www9.informatik.uni-erlangen.de/External/vollib/}},
a Jet simulation dataset, and an archaeological scan of \emph{Spathorhynchus fossorium}~\footnote{\url{http://www.digimorph.org}}. We use
three different types of volume rendering models. We consider no illumination, corresponding to the basic emission-absorption model
of Equation~\ref{eq:vri}. We also use OSPRay~\cite{wald2017ospray} to generate images under direct illumination, as well as global illumination
effects. In particular, we use volumetric ambient occlusion with $128$ samples, and $8$ samples per pixel, and use an HPC cluster to accelerate image generation time.
We use a fixed directional light source
for illumination, defined relative to the viewpoint. Table~\ref{table:datasets} (left) summarizes dataset statistics and lighting models used for each dataset,
while Table~\ref{table:model} lists the size as well as timings of our network for training and the different applications. Note that these values
are independent of dataset.}

\subsection{TF Sensitivity}

\begin{table}[!b]
\begin{center}
   \begin{tabular}{| c | c | c | c | c |}
      \hline
        Size & Train & Render & TF Explore & TF Sensitivity \\ \hline
		101 MB & 16.5 hr & .007 s & .06 s & .49 s \\ \hline
   \end{tabular}
\end{center}
\caption{We list the size of our network, and timings for training, rendering an image, TF exploration, and TF sensitivity.}
\label{table:model}
\end{table}

\begin{figure*}[!t]
	\begin{center}
	\includegraphics[width=0.9\linewidth]{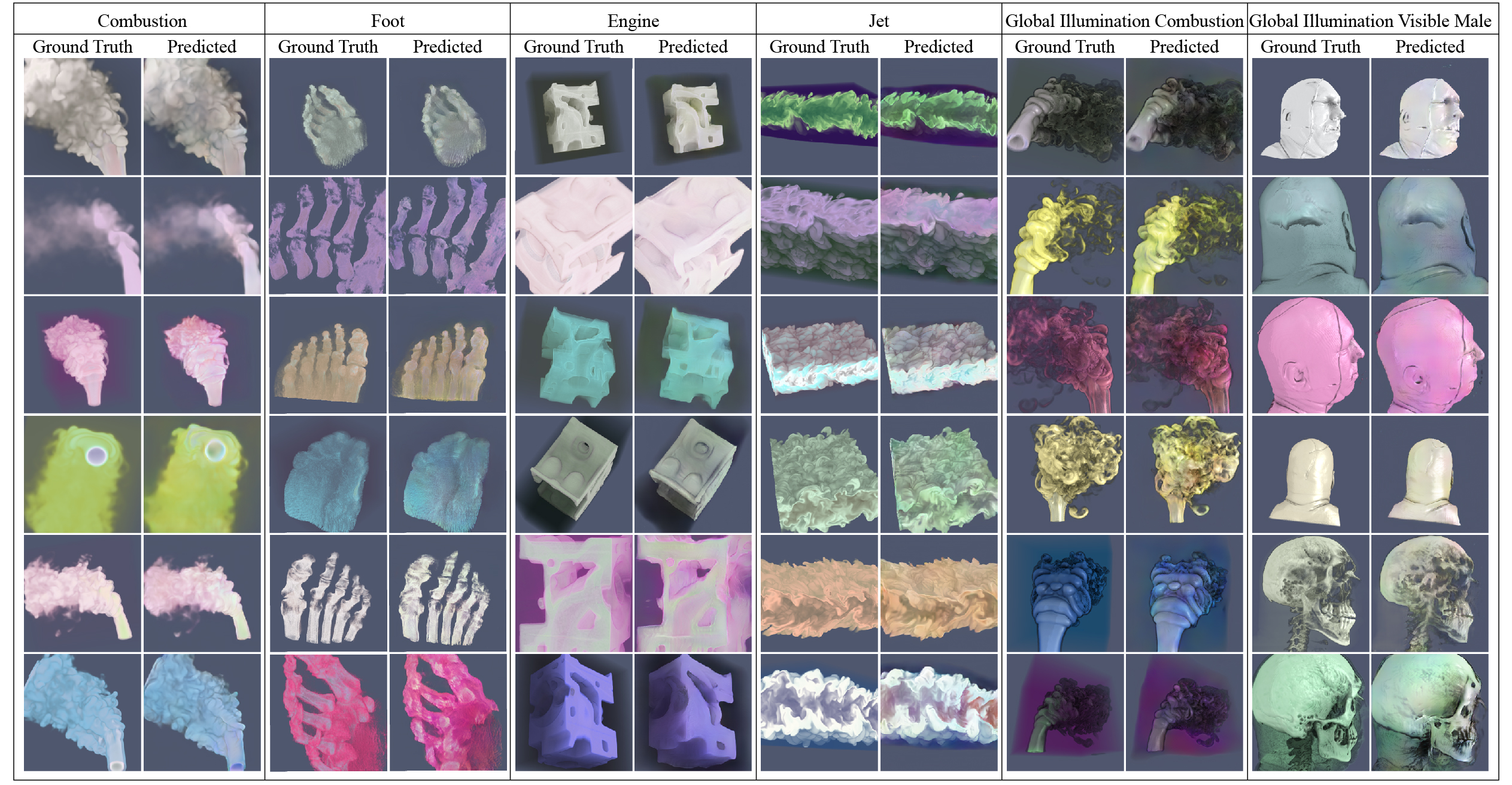}
		\caption{We show qualitative results comparing synthesized images to ground truth volume renderings produced without illumination.
		The bottom row shows typical artifacts, such as incorrect color mapping and lack of detail preservation.}
	\label{fig:qualitative}
	\end{center}
\end{figure*}

\revision{We show how to use TF sensitivity to guide the user in making impactful TF edits. Fig.~\ref{fig:sense_region} depicts
a typical interaction scenario for Region Sensitivity Plots for the Combustion volume with direct illumination.
The user first selects a region (top), here shown as a slightly transparent white rectangle, and we compute the region's sensitivity plot,
shown as the red plot on the right. High values suggest portions of the TF domain that, upon a user edit, will result in a
change in the image. By adding a small mode to the opacity TF GMM, we can observe (mid-left) that this portion of the TF domain
corresponds to the primary flame of the plume. Subsequently selecting the base of the plume, we update the sensitivity
plot (mid-right). By adding a mode to a sensitive region, we see (bottom-left) that this resulted in higher density covering the base,
with the white material being covered by the purple material.}

\revision{We next show usage of Scalar Value Sensitivity Fields for understanding how modifications to a portion of the TF domain
can impact image regions. We apply this on the Foot dataset in Fig.~\ref{fig:sense_field}. The upper left image
corresponds to the TF shown on the right. In the middle we show the sensitivity field corresponding to the shaded red
region selected on the TF. We observe that locations of high sensitivity exist along the bone of the foot. By adding
a mode to the TF at this scalar value, we observe (middle-left) that indeed this value corresponds to an increase in the bone
density. Subsequently selecting a region of the TF (middle-right) updates the field (middle-left), with more of the bone portions
of the foot highlighted. Adding a mode to the TF at this value shows that this edit fills in the bone, in addition to part
of the ambient volume (lower-left). Note that the ambient volume did not change as much as the bone of the foot, as suggested
by the sensitivity field. For this example, we stress that the field sensitivity is small relative to the global sensitivity,
as we visually encode the field based on the user selection through opacity.}

\subsection{Opacity TF Exploration}

\revision{We next show an example of volume rendering exploration using the opacity TF latent space. We study opacity TF variation for
the Jet dataset, see Fig.~\ref{fig:exploration}. This dataset corresponds to a simulation of jet flames,
where the scalar field is taken to be the mixture fraction. Here the user first selects most of the t-SNE projected latent space (left). This provides
a general overview of the dataset, where we can observe a low mixing with fuel in the upper right
portion of the projection space, and a progressively larger mixture fraction towards the bottom left. The user also hovers over a portion
of the latent space projection, shown as a Gaussian blob in dark blue, to synthesize an image shown in the top-right.
Upon decoding from the opacity TF latent space we see that the reconstructed TF has low opacity value near
the high mixture ratio, namely it trails off after 0.5. This is consistent with the shown image which has little material in the middle of the volume.}

\revision{The user then changes their view to the other side of the volume, zooms in, and selects a smaller portion of the projected latent space (right).
The more refined selection results in finer-grained shape variations throughout the volume. The user's selection of the latent space,
corresponding to the upper-right image, indicates higher fuel mixing compared with that in (a). The reconstructed TF further corraborates
this, as we see a larger TF value being assigned to a higher mixture ratio relative to (a).}

\revision{Our TF exploration interface also enables the user to better understand the relationship between features in the volume and the
corresponding relevant domain of the opacity TF. In Fig.~\ref{fig:tf_interpolate} we show the Combustion dataset for OSPRay-rendered
images at four different user selections in the opacity TF latent space. In the first three images we observe two primary modes
in the TF, where by browsing the latent space the user can observe how the TF changes. It becomes clear that the mode on the left,
i.e. low scalar values, corresponds to the flame of the plume, while the right mode impacts the handle.}

\subsection{Model Validation}

\revision{We validate our model through evaluation on a hold-out set of $2,000$ volume-rendered images, or images not used to train
the network, to assess model generalization for each dataset. We use Root Mean Squared Error (RMSE) as an evaluation metric. RMSE alone, however, may
fail to capture other useful statistics in the output, and is sensitive to errors in pose. Hence, to measure higher-level similarity we
compute distance between color histograms with the Earth Mover's Distance (EMD). EMD helps mitigate misalignment in the
histogram space~\cite{pele2009fast}. The cost between histogram bins is normalized such that the maximum cost is $1$.}


\revision{Table~\ref{table:datasets} (right) reports evaluation in terms of the mean RMSE and EMD for all datasets.
Overall we find the Image RMSE and Color EMD to be within a reasonable error tolerance, though we can observe several trends
based on the datasets. First, error tends to increase with the use of more advanced illumination models. Secondly, we observe
that as the volume resolution increases, the error also increases. Both of these data characteristics are likely to contribute to
a larger number of features present in the volume rendered images, and learning these features can pose a challenge for our network.}

We show qualitative results for volumes rendered without illumination in the first four columns of Fig.~\ref{fig:qualitative}.
We find that our architecture is quite effective in synthesizing pose and shape, suggesting that our opacity GAN is effective at capturing
coarse information, while the translation GAN is effective in using the opacity image to synthesize more detailed features. Nevertheless,
the translation GAN may not always succeed in producing the right color mapping. We show such typical artifacts in the right column
for Combustion, Foot, and Jet. \revision{Furthermore, we also show an artifact in the opacity TF for the Engine dataset in failing to preserve the hole
in the center-left of the image.}

\revision{The last two columns of Fig.~\ref{fig:qualitative} show results for volumes rendered with global illumination. Note that our model is effective
at capturing various shading effects -- specular highlights, self-shadowing -- in addition to the details present in the volume.
Nevertheless, we can observe in Table~\ref{table:datasets} that the RMSE does increase when using global illumination compared to
volumes rendered without illumination. However we are still able to capture the color distribution, as indicated by the EMD,
with a global illumination model. We generally find similar artifacts to those images rendered without illumination, as shown by
the incorrect color mapping in Combustion, and incorrect shape inferred by the opacity TF in the Visible Male dataset.
We also observe small skull details not preserved in the fifth row for Visible Male.}

\subsubsection{Baseline Comparisons}
\label{subsec:baseline}

\revision{To validate our approach and network design choices we have compared our approach to several baselines. First, we would like to verify that our network is not overfitting, i.e. simply
memorizing images in the training dataset. Thus we consider a nearest-neighbor baseline, where given a ground-truth image we find its nearest image in the training dataset.
For efficient search we use the hashing-based retrieval approach of Min et al.~\cite{min2010compact}. For our second comparison we would like to verify how significant
the adversarial loss is relative to a purely image-based loss. To this end, we modify the translation GAN generator such that it is replace by an image-based $l_1$ loss,
namely the adversarial loss in Equation~\ref{eq:translationgan} is removed. Conversely, for our third comparison we would like to verify the impact of removing
the image-based $l_1$ loss from Equation~\ref{eq:translationgan}, thus we only optimize for the adversarial loss.}

\revision{Table~\ref{table:baseline} shows the mean RMSE and EMD for all baselines evaluated on the Combustion dataset with direct illuination, with our proposed approach denoted GAN+$l_1$.
We observe that the Image RMSE for the nearest neighbor baseline (NN) is comparable if slightly better than our approach, but the Color EMD is worse.
This suggests that our approach is able to synthesize color details not present in the training data via the learned color mapping of our network.
Similar observations can be made in comparing the adversarial-only loss (GAN) with our approach, which shows the benefit of adding an image-based $l_1$ loss to aid in the color mapping.
Using only an $l_1$ loss, on the other hand, produces a much lower Image RMSE and slightly lower Color EMD. A smaller Image RMSE is expected in this setting,
since the objective function and error measure are both image-based, whereas the adversarial loss is not. Namely, the generator in a GAN never directly observes
images in the training dataset, its updates are solely based on the discriminator's model of real/fake images.}

\begin{table}[!t]
\begin{center}
   \begin{tabular}{| c | c | c | c | c |}
      \hline
      Evaluation & NN & $l_1$ & GAN & GAN+$l_1$ \\ \hline
      Image RMSE & 0.059 & 0.047 & 0.060 & 0.060 \\ \hline
      Color EMD & 0.020 & 0.007 & 0.017 & 0.011 \\ \hline
   \end{tabular}
\end{center}
\caption{We show quantitative results for our method compared to baselines of nearest neighbor retrieval, $l_1$ loss, and GAN loss.}
\label{table:baseline}
\end{table}

\begin{figure}[!t]
	\begin{center}
	\includegraphics[width=0.9\linewidth]{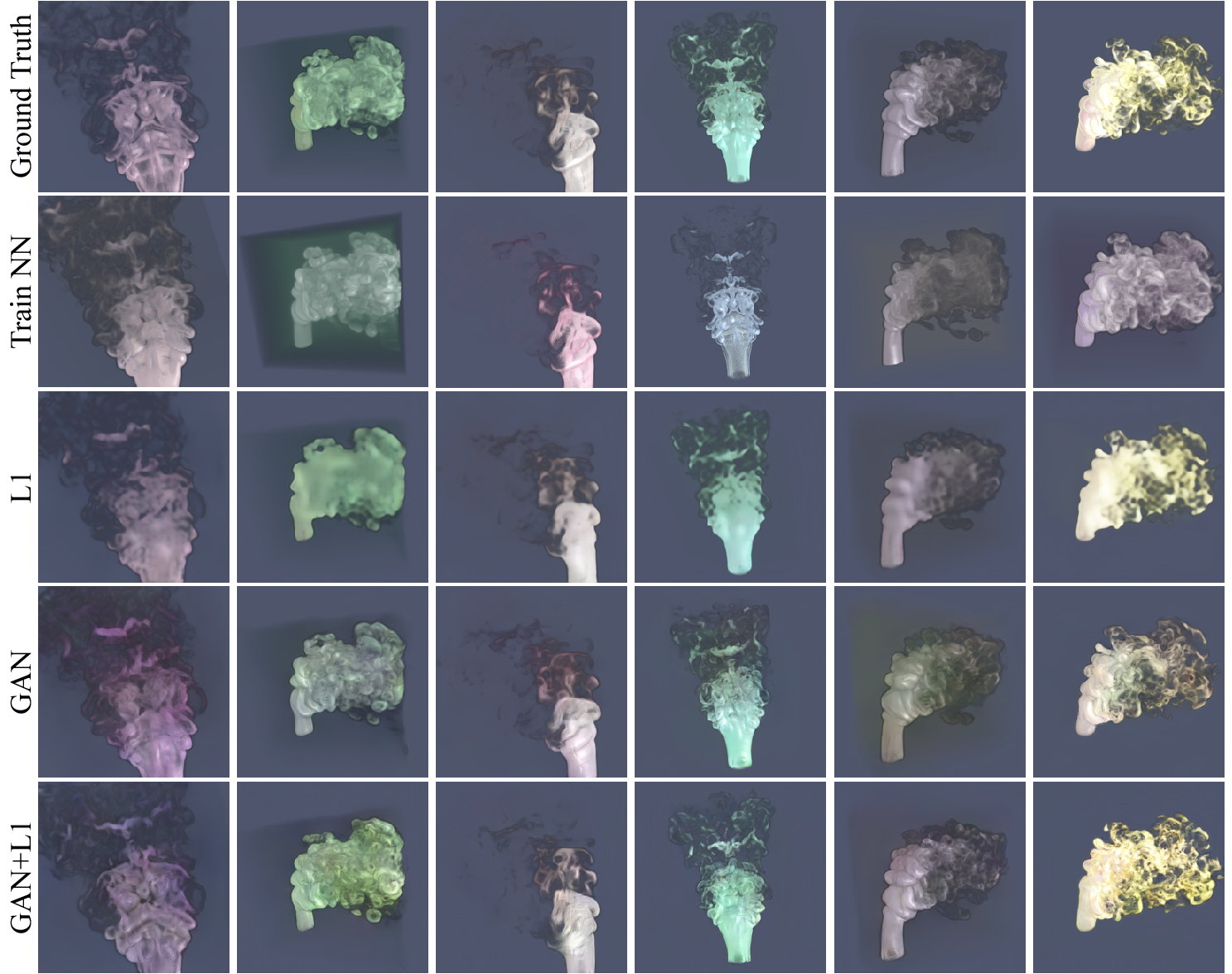}
		\caption{We compare our method to training dataset nearest neighbor retrieval, image-based $l_1$ loss, and GAN loss. Nearest neighbor tends to incorrectly predict color,
		the $l_1$ loss blurs details, and the GAN loss can result in color shifts. GAN+L1 strikes a balance between preserving detail and color.}
	\label{fig:baseline}
	\end{center}
\end{figure}

\revision{Fig.~\ref{fig:baseline} shows qualitative results for the baselines. We find the nearest-neighbor approach is effective at retrieving similar poses,
but the color and opacity are not necessarily preserved. This is the cause for the competitive Image RMSE but smaller Color EMD, as small perturbations in pose can
result in large RMSE error. The $l_1$ loss is effective at preserving color, but is unable to reproduce fine details compared to using an adversarial loss. This
is the primary issue with solely using an image-based loss, as although the reported Image RMSE is low, details become blurred out, as other works have
identified~\cite{pathak2016context,Isola_2017_CVPR}. The adversarial-only loss is capable of reproducing details, but there exists small color shifts
in the generated images. Our proposed approach strikes the best balance in generating details through the adversarial loss, while preserving color
through the $l_1$ loss.}

\subsubsection{Opacity TF Latent Space Dimensionality}

\revision{We validate our choice of opacity TF latent space dimension, as discussed in Sec.~\ref{subsec:opacitygan}, by comparing networks with different dimensionalities,
namely 4, 8, 16, 32, and 64. In order to reduce the large computational burden of training all networks
we modify our architecture to produce $64 \times 64$ resolution images by removing the last few upsampling layers, analagous to
the evaluation performed in Dosovitskiy et al.~\cite{dosovitskiy2015learning}. We set $\lambda = 0$ in Equation~\ref{eq:translationgan}
in order to remove the influence of the $l_1$ loss, since the opacity TF latent space largely impacts shape and not color.}

\begin{figure}[!t]
	\begin{center}
	\begin{subfigure}[b]{0.14\textwidth}
		\includegraphics[width=\linewidth]{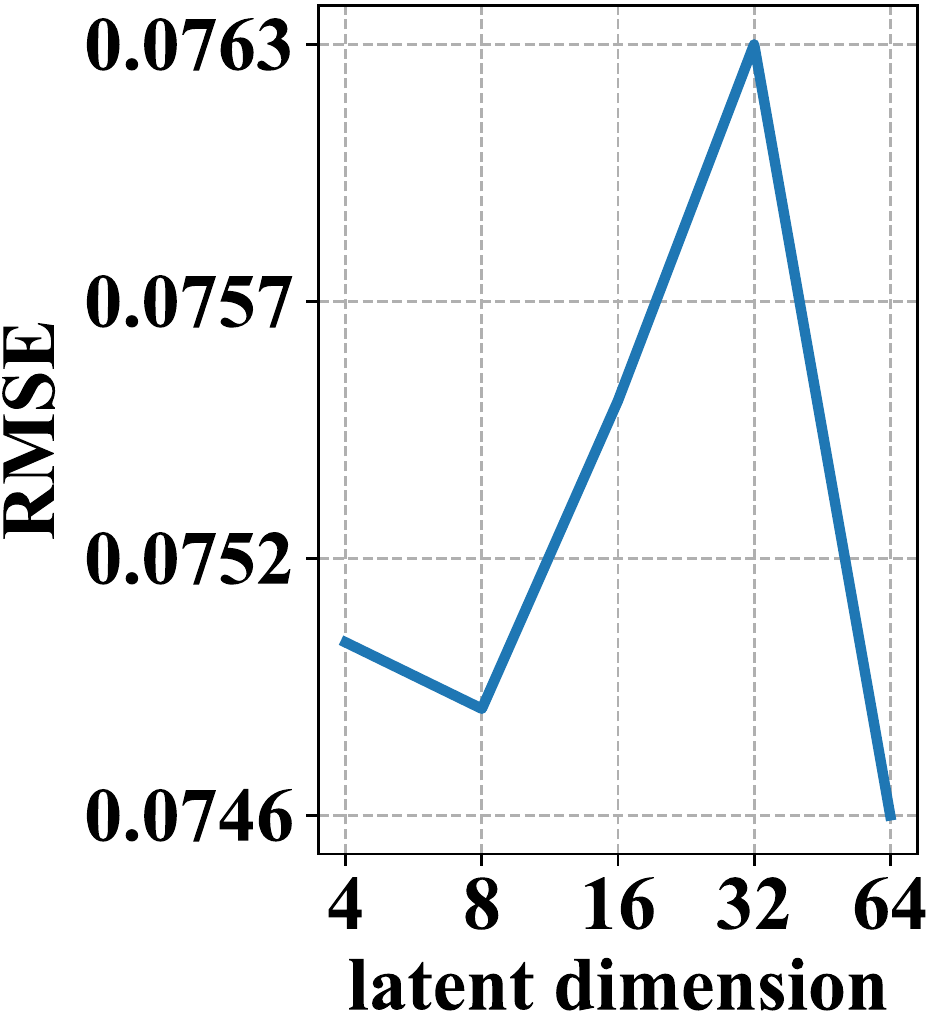}
		\caption{Opacity Image}
		\label{subfig:latent-op-rmse}
	\end{subfigure}
	\begin{subfigure}[b]{0.14\textwidth}
		\includegraphics[width=\linewidth]{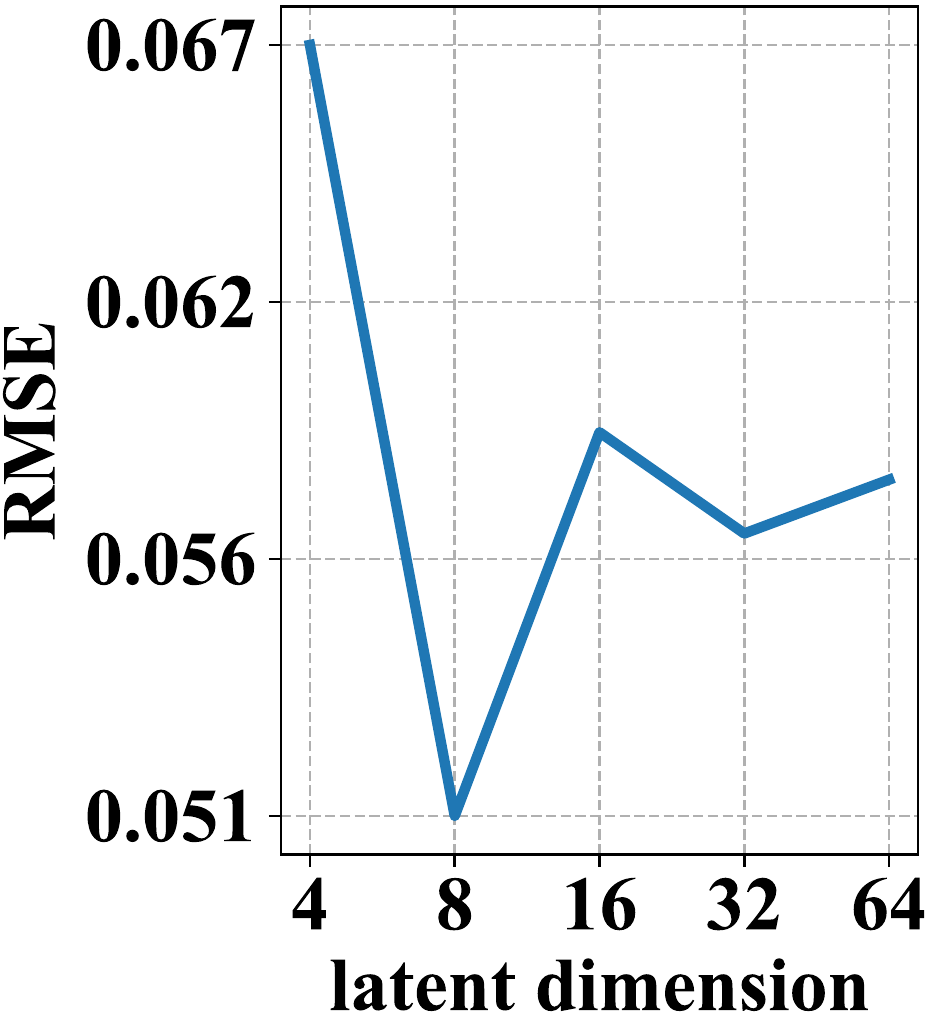}
		\caption{Color Image}
		\label{subfig:latent-rgb-rmse}
	\end{subfigure}
	\begin{subfigure}[b]{0.14\textwidth}
		\includegraphics[width=\linewidth]{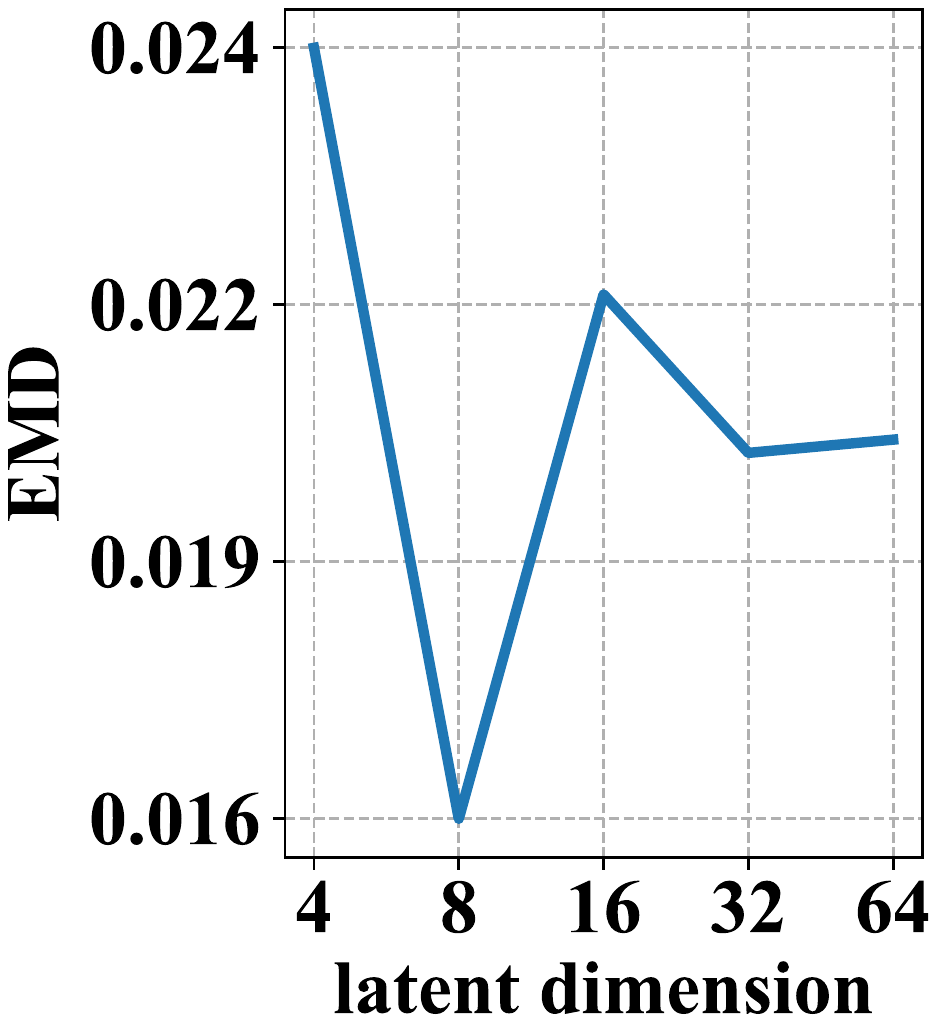}
		\caption{Color Histogram}
		\label{subfig:latent-rgb-emd}
	\end{subfigure}
    \end{center}
	 \caption{We evaluate varying the dimensionality of the opacity TF latent space for Combustion. Although the opacity errors are small, we observe
	 larger error variation in the color image. The results suggest a dimension of 8 is best.}
	 \label{fig:latent}
\end{figure}

\revision{We have evaluated the networks on the Combustion dataset, using no illumination.
Fig.~\ref{fig:latent} shows error plots for the opacity image RMSE (a), color image RMSE (b), and color histogram EMD (c).
We find that the latent space dimensionality does not much impact the quality of the opacity images, but there exists more significant
differences in the color images. We see that a latent dimension of $8$ performs best for this experiment. Though one might expect a
larger dimension to perform better, in general the dimension should set such that the latent space captures the primary shape variations throughout the
volume, and overestimating this dimension could result in poorer generalization. We have thus used $8$ throughout all of our experiments.}

\revision{We acknowledge that a dimension of $8$ may not be ideal for all other volumes. For the datasets we have considered we found this to
work reasonably well, but for more complex datasets cross-validation can be performed to optimize the latent dimensionality. Nevertheless,
high-dimensional latent spaces (i.e. $\gg 8$) can have an impact on the exploration of the TF latent space. In particular, a high-dimensional
space is more difficult to sample in generating a set of TFs, as discussed in Sec.~\ref{sec:tffeats}. Thus we see a trade-off between image quality
and downstream applications of the network, which is ultimately a user choice depending on their needs.}

\subsubsection{Influence of $l_1$ Loss}

\revision{In Sec.~\ref{subsec:baseline} we showed how the combination of the adversarial loss and the $l_1$ retained feature details and preserved color,
respectively, with the $l_1$ loss contribution $\lambda$ set to $150$. We now study the setting of $\lambda$, where we consider values of 50, 150, and 450.
We experimentally verified that these values correspond to the $l_1$ loss contribution being $\frac{1}{3}$, 1, and $3\times$ the amount of the adversarial loss,
respectively, though it is challenging to precisely set $\lambda$ relative to the adversarial loss due to the dynamics of training GANs~\cite{arjovsky2017wasserstein}.}

\begin{figure}[!t]
	\begin{center}
	\includegraphics[width=0.85\linewidth]{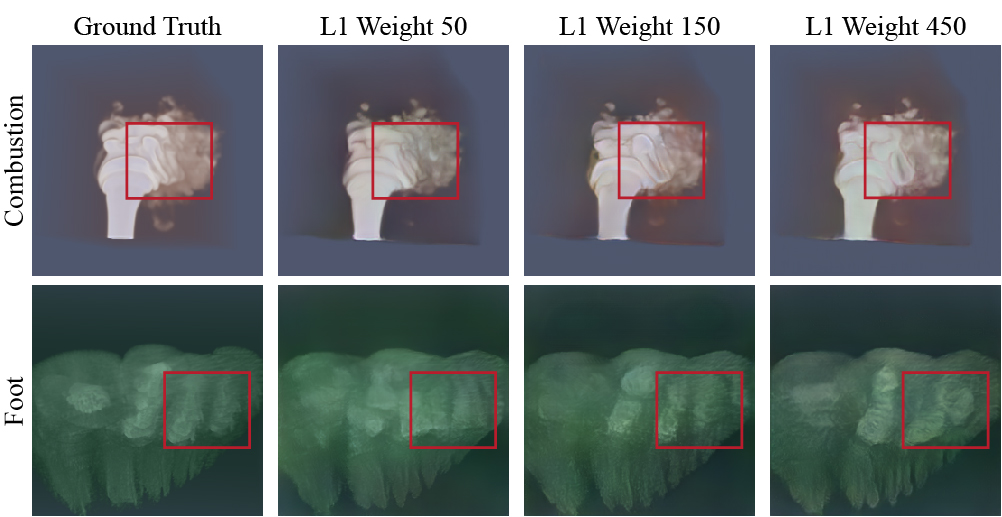}
		\caption{We compare results in varying the weight of the $l_1$ loss. In certain cases a large weight may fail to
		preserve detail, while a small weight results in color shift, as shown in Combustion.}
	\label{fig:l1-qualitative}
	\end{center}
\end{figure}

\revision{We have trained networks for the Combustion and Foot datasets without illumination, synthesizing images of $256 \times 256$.
Fig.~\ref{table:l1} summarizes the results, showing the mean Image RMSE and Color EMD error metrics. In general, we can observe that the error measures
decrease as $\lambda$ increases, though overall the differences are not too significant, particulary for the Foot dataset. Qualitative results in
Fig.~\ref{fig:l1-qualitative} show that $\lambda=450$ may fail to preserve the highlighted details, while for $\lambda=50$ we can observe a
color shift in the Combustion example. Thus $\lambda=150$ strikes a compromise between detail and color, though the results indicate that
the network quality does not change too much for the given range of $\lambda$, showing that this parameter is fairly insensitive to set.}

\begin{table}[!t]
\begin{center}
   \begin{tabular}{| c | c | c | c | c |}
      \hline
      Dataset & Evaluation & $l_1$=50 & $l_1$=150 & $l_1$=450 \\ \hline
      Combustion & Image RMSE & 0.050 & 0.046 & 0.044 \\ \hline
      Combustion & Color EMD & 0.015 & 0.011 & 0.012 \\ \hline
      Foot & Image RMSE & 0.065 & 0.064 & 0.062 \\ \hline
      Foot & Color EMD & 0.019 & 0.017 & 0.016 \\ \hline
   \end{tabular}
\end{center}
\caption{We compare the setting of the $l_1$ loss in the optimization for different weights. Generally, we see that larger weights result
	in lower Image RMSE, but for weights of 150 and 450 the color distributions are fairly similar.}
\label{table:l1}
\end{table}

\section{Discussion}

Generative models provide a unique perspective on the process of volume rendering, and we see a number of directions to take in
improving our approach and adapting it for other uses. A limitation is the time required for training,
particularly the translation GAN, which requires 16.5 hours to train. Deep learning is, however, quite amenable to data parallelism
since gradients are computed on minibatches, hence training could be accelerated given multiple GPUs. Furthermore, in large-scale
numerical simulations, computation times can easily be comparable to our training times, hence one potential application
is to train our network in-situ, as volumetric scalar data is produced by a simulation. This setup also suggests the possibility to design
a network to learn from both time-varying and multivariate data. As there likely exists
significant structure/correlation between these types of data, a single network should be capable of learning from these forms of data as
they are output by a simulation, providing a significant form of data compression for the purposes of offline rendering.

\revision{Although our model incurs errors within a reasonable tolerance, we think that there exists opportunities
to improve the quality of the results. Currently we condition on opacity to help stabilize training, however a limitation
of opacity is that it can saturate, providing very coarse shape cues. We think depth-based measurements can be computed to provide
better information, while still being reasonable to predict. We also think that alternative network architectures that better align
the color and opacity TF can be developed to improve on our current limitations in color mapping.}

Note that in our learning setup we have full control over the training data that is generated. In our approach
we make as few assumptions on the data as possible in generating random viewpoints and TFs. A disadvantage
with this approach, however, is that certain views or TFs may be poorly sampled, and thus generalization will suffer.
\revision{It is worth exploring different ways of sampling views and TFs that improve generalization, perhaps in a data-driven manner
where views and TFs that incur high error are adaptively sampled. An approach that generates data during training could also
help in optimizing the amount of data necessary, which as shown can be an overhead as large as training depending
on the illumination model and volume.}

To make our model more practical we need to consider other forms of volume interaction.
\revision{For instance, volume clipping and lighting modifications are two common parameters in volume interaction,
and we think its possible to encode both as additional visualization parameters in our model.}
Furthermore, 1D TFs are widely recognized as having
limitations in identifying volumetric features. The incorporation of various forms of 2D TFs
into our model should require little modification, effectively replacing 1D convolutions with 2D. We intend to explore
how different types of TFs can benefit our model, potentially leading to novel ways for TF exploration,
similar to our opacity TF latent space.

Our approach is designed to analyze a single volumetric dataset, however we think there are interesting research directions
to take for GANs in conditioning on the volume too.
This could lead to novel ways of synthesizing volume-rendered images from volumes not seen at training time. Alternatively,
one could consider GANs for synthesizing TFs, rather than images, conditioned on a given volume.
More generally we think generative models, can provide a host of novel ways to interact with volumetric data.

\section*{Acknowledgements}

We thank Peer-Timo Bremer for stimulating discussions. This work was partially supported by the National Science Foundation IIS-1654221.

\bibliographystyle{IEEEtran}
\bibliography{IEEEabrv,paper}

\end{document}